\documentclass[10pt]{article}
\usepackage{amsmath}
\usepackage{graphicx}
\usepackage{amsfonts}
\usepackage{amssymb}
\usepackage{epsf}
\usepackage{latexsym}

\def\80{\hspace{0.8in}}

\newcommand{\be}{\begin{enumerate}}
\newcommand{\ee}{\end{enumerate}}
\newcommand{\bi}{\begin{itemize}}
\newcommand{\ei}{\end{itemize}}
\newcommand{\bd}{\begin{description}}
\newcommand{\ed}{\end{description}}

\def\beq{\begin{equation}}
\def\eeq{\end{equation}}
\def\bea{\begin{eqnarray}}
\def\eea{\end{eqnarray}}
\def\hat{\widehat}

\def\pa{\partial}
\def\d{\textrm{d}}

\def\ttH{\mbox{\tt H}}

\def\brho{\mbox{\boldmath$\rho$}}
\def\cr{\mbox{\scriptsize{\bf $\mbox{ } \times \mbox{ }$}}}

\def\ma{\mbox{a}}
\def\mb{\mbox{b}}

\def\md{\mbox{d}}

\def\mh{\mbox{h}}

\def\ml{\mbox{l}}

\def\mn{\mbox{n}}

\def\mq{\mbox{q}}

\def\mC{\mbox{C}}
\def\mD{\mbox{D}}

\def\mF{\mbox{F}}

\def\mH{\mbox{H}} 

\def\mJ{\mbox{J}}

\def\mL{\mbox{L}}

\def\mN{\mbox{N}} 

\def\mP{\mbox{P}}

\def\mR{\mbox{R}}

\def\sa{\mbox{\scriptsize a}}
\def\sb{\mbox{\scriptsize b}}
\def\sc{\mbox{\scriptsize c}}
\def\sd{\mbox{\scriptsize d}}
\def\se{\mbox{\scriptsize e}}

\def\sh{\mbox{\scriptsize h}} 
\def\si{\mbox{\scriptsize i}}

\def\sm{\mbox{\scriptsize m}}
\def\sn{\mbox{\scriptsize n}} 
\def\so{\mbox{\scriptsize o}}

\def\sr{\mbox{\scriptsize r}}
\def\sss{\mbox{\scriptsize s}}
\def\st{\mbox{\scriptsize t}}
\def\su{\mbox{\scriptsize u}}

\def\sw{\mbox{\scriptsize w}}
\def\sx{\mbox{\scriptsize x}}

\def\sA{\mbox{\scriptsize A}}

\def\sD{\mbox{\scriptsize D}}

\def\sF{\mbox{\scriptsize F}}
\def\sG{\mbox{\scriptsize G}}
\def\sH{\mbox{\scriptsize H}}
\def\sI{\mbox{\scriptsize I}}

\def\sK{\mbox{\scriptsize K}}

\def\sN{\mbox{\scriptsize N}} 
\def\sO{\mbox{\scriptsize O}}

\def\sR{\mbox{\scriptsize R}}

\def\sT{\mbox{\scriptsize T}}

\def\eph(B){\mbox{\scriptsize emergent(LMB)}}

\def\ta{\mbox{\tiny a}}

\def\tc{\mbox{\tiny c}}
\def\td{\mbox{\tiny d}}
\def\te{\mbox{\tiny e}}

\def\th{\mbox{\tiny h}}
\def\ti{\mbox{\tiny i}}

\def\tl{\mbox{\tiny l}}
\def\tm{\mbox{\tiny m}}
\def\tn{\mbox{\tiny n}}
\def\to{\mbox{\tiny o}}

\def\tr{\mbox{\tiny r}}

\def\ttt{\mbox{\tiny t}}

\def\tD{\mbox{\tiny D}}

\def\tN{\mbox{\tiny N}}

\def\tT{\mbox{\tiny T}}

\def\sbP{\mbox{{\bf \scriptsize P}}}

\def\sbR{\mbox{{\bf \scriptsize R}}}

\def\fH{\mbox{\sffamily H}}

\def\fP{\mbox{\sffamily P}}
\def\fQ{\mbox{\sffamily Q}}
\def\fR{\mbox{\sffamily R}}
\def\fS{\mbox{\sffamily S}}
\def\fT{\mbox{\sffamily T}}

\begin{document}
\begin{titlepage}
\vspace{.7in}
\begin{center}
 
\vspace{2in} 

{\LARGE\bf QUANTUM COSMOLOGICAL RELATIONAL MODEL} 

\vspace{.1in}

{\LARGE\bf OF SHAPE AND SCALE IN 1-$d$ }

\vspace{.2in}

{\bf Edward Anderson}$^{1}$ 

\vspace{.2in}

{\em $^1$ DAMTP Cambridge U.K.}

\vspace{.2in}

\end{center}

\begin{abstract}

Relational particle models are useful toy models for quantum cosmology and the problem of time 
in quantum general relativity.
This paper shows how to extend existing work on concrete examples of relational particle models 
in 1-d to include a notion of scale. 
This is useful as regards forming a tight analogy with quantum cosmology and 
the emergent semiclassical time and hidden time approaches to the problem of time.  
This paper shows furthermore that the correspondence between relational particle models and classical 
and quantum cosmology can be strengthened using judicious choices of the mechanical potential.
This gives relational particle mechanics models with analogues of spatial curvature, 
cosmological constant, dust and radiation terms. 
A number of these models are then tractable at the quantum level.  
These models can be used to study important issues 1) in canonical quantum gravity: the problem of time, 
the semiclassical approach to it and timeless approaches to it (such as the na\"{\i}ve Schr\"{o}dinger 
interpretation and records theory).
2) In quantum cosmology, such as in the investigation of uniform states, robustness, and the qualitative 
understanding of the origin of structure formation.     

\end{abstract}

\vspace{0.2in}


\vspace{0.2in}

\mbox{ }

\noindent$^1$ ea212@cam.ac.uk  

\mbox{ }

\end{titlepage}

\section{Introduction}

\subsection{What are relational particle models (RPM's)?}

Scaled relational particle mechanics (RPM) (originally proposed in \cite{BB82} and further studied in 
\cite{ERPM, B94I, EOT, 06I, TriCl, 08I, Cones} is a mechanics in which only relative times, relative angles 
and relative separations have physical meaning.  
On the other hand, pure-shape RPM (originally proposed in \cite{B03} and further studied in \cite{SRPM, 
06II, TriCl, FORD, 08I, 08II, +tri, AF, FileR}) is a mechanics in which only relative times, relative 
angles and ratios of relative separations have physical meaning.  
More precisely, these theories implement the following two Barbour-type (Machian) 
relational\footnote{RPM's are 
relational in Barbour's sense of the word rather than Rovelli's distinct one. 
\cite{Rovellibook, B94I, EOT} are original references for these authors' original material, while 
\cite{08I} discusses some of the differences between them.} 
postulates. 

\mbox{ } 

\noindent 1) They are {\it temporally relational} \cite{BB82, RWR, Lan, FORD}, i.e. there is no 
meaningful primary notion of time for the whole system thereby described (e.g. the universe). 
This is implemented by using actions that are manifestly reparametrization invariant while also being 
free of extraneous time-related variables [such as Newtonian time or the lapse in General Relativity 
(GR)].   
This reparametrization invariance then directly produces primary constraints quadratic in the momenta 
\cite{Dirac}. 

\mbox{ } 

\noindent 2) They are {\it configurationally relational}, which can be thought of in terms of a certain 
group $G$ of transformations that act on the theory's configuration space $\fQ$ being held to be 
physically meaningless \cite{BB82, RWR, Lan, FORD, FEPI, Cones}.   
One implementation of this uses arbitrary-$G$-frame-corrected quantities rather than `bare' 
$\fQ$-configurations.
For, despite this augmenting $\fQ$ to the principal bundle $P(\fQ, G)$, variation with respect to each 
adjoined independent auxiliary $G$-variable produces a secondary constraint linear in the momenta which 
removes one $G$ degree of freedom and one redundant degree of freedom from $\fQ$.   
Thus, one ends up on the desired reduced configuration space -- the quotient space $\fQ/G$.  
Configurational relationalism includes as subcases both spatial relationalism (for spatial 
transformations) and internal relationalism (in the sense of gauge theory).  
For scaled RPM, $G$ is the Euclidean group of translations and rotations, while for pure-shape RPM it 
is the similarity group of translations, rotations and dilations.

\subsection{Motivation for RPM's: Toy models for classical and quantum GR}

My principal motivation for studying RPM's\footnote{RPM's have elsewhere been motivated by the 
long-standing absolute or relational motion debate, and by RPM's making useful examples in the study of 
quantization techniques \cite{BS89etc, Banal, FileR}. 
The present paper's motivation follows from that in \cite{K92}. 
Moreover, I have now considerably expanded on this motivation by providing a very large number of 
analogies between RPM's and Problem of Time strategies. 
(See \cite{Cones} for a more detailed account.)}
is their utility as toy model analogues of GR in its traditional dynamical form (`geometrodynamics':  
the evolution of spatial geometries).  
The extent of the analogies between RPM's and GR (particularly in the formulations \cite{BSW, RWR, 
ABFKO} of geometrodynamics) is comparable but different to the resemblance between GR and the more 
habitually studied minisuperspace models \cite{Mini, HH83, Wiltshire}.  
{\sl RPM's are likely to be comparably useful as minisuperspace from the perspective of theoretical toy 
models}.
Some principal RPM--GR analogies are 

\noindent
1) that the quadratic energy constraint\footnote{I use
$a$, $b$, $c$ as particle label indices running from 1 to $N$ for particle positions, 
$e$, $f$, $g$ as particle label indices running from 1 to $n = N - 1$ for relative position variables,  
$i$, $j$, $k$ as spatial indices,
$p$, $q$, $r$ as relational space indices (in 1-$d$, these run from 1 to $n$ and are interchangeable 
with $e$, $f$, $g$), and
$u$, $v$, $w$ as shape space indices (in 1-$d$, these run form 1 to $n$ -- 1 = $N$ -- 2).
[I reserve $d$ to denote dimension, $n$ for the number of relative position variables, 
$h$ for heavy and $l$ for light.]
I also use straight indices (upper or lower case) to denote quantum numbers, the index $S$ to denote 
`shape part' and the index $\rho$ (referring to the hyperradius) to denote `scale part'.
The formulation of the RPM that I present here is in relative Jacobi coordinates $\sbR^e$ (see Sec 2 for 
a detailed explanation of what these are), the conjugate momenta to which are the $\sbP_e$.  
$N^{ef ij} = \delta^{ef}\delta^{ij}/\mu_i$ is the inverse of the kinetic metric 
on the corresponding configuration space, where the $\mu_i$ are the particle cluster masses associated 
with the Jacobi coordinates.
$h_{\alpha\beta}$ is the spatial 3-metric, with determinant $h$, covariant derivative $D_{i}$, 
Ricci scalar $R$ and conjugate momentum $\pi^{ij}$.
${\cal N}_{ijkl} =\{h_{ik}h_{jl} - h_{ij}h_{kl}/2\}/\sqrt{h}$ is the DeWitt supermetric of GR.}
\beq 
\ttH := N^{ef ij}\mP_{e i}\mP_{f j}/2 + V = E 
\eeq 
is the analogue of GR's quadratic Hamiltonian constraint,
\beq
{\cal H} := {\cal N}_{ijkl}\pi^{ij}\pi^{kl} - \sqrt{h}R = 0  
\label{GRham}
\eeq 

\noindent 2) That RPM's linear zero total angular momentum constraint 
\beq
{\bf L} := \sum\mbox{}_{\mbox{}_{\mbox{\scriptsize $e$}}} {\bf R}_e \cr {\bf P}_e = 0 
\label{ZAM}
\eeq 
is a nontrivial analogue of GR's linear momentum constraint 
\beq
{\cal L}_{i} := - 2D_{j}\pi^{j}\mbox{}_{i} = 0 \mbox{ } . 
\label{GRmom}
\eeq
3) There are analogies between RPM's and GR at the level of the configuration spaces involved (see Sec 2).  

\noindent
4) There are analogies between RPM's and GR at the level of the actions involved (for 3 + 1 
geometrodynamical formulations of GR, see Sec 3).

\noindent
5) In GR, 2) and the notion of local structure/clustering are tightly related as both concern the 
nontriviality of the spatial derivative operator.  
However, for RPM's, the nontriviality of angular momenta and the notion of 
structure/inhomogeneity/ particles clumping together are unrelated.  
Thus, even in the simpler case of 1-$d$ models considered in the present paper, RPM's have nontrivial 
notions of structure formation/inhomogeneity/localization/correlations between localized quantities.  
In the subsequent 2-$d$ model paper \cite{08III}, one has nontrivial linear constraints as well.
Each of these features is important for many detailed investigations in Quantum Gravity and Quantum 
Cosmology.

2) and 5) are specific ways in which RPM's are more useful than minisuperspace models (linear 
constraints and inhomogeneity/clustering are trivial for minisuperspace via all points in space 
being equivalent in minisuperspace models).  
On the other hand, minisuperspace possesses an indefinite kinetic term like GR does, and a 
GR-like potential term.  
(RPM's, being mechanics, have a positive-definite kinetic term and largely unrestricted 
potentials, which potential freedom is, however, usefully exploited in the present paper 
to further align specific RPM's with GR quantum cosmological models.)  
Other similarly complex toy models of gravitation are 
i) 2 + 1 GR \cite{Carlip}, which is still finite (it possesses global degree of freedom only); this comes 
with the advantages of a nontrivial notion of diffeomophism 
and some 3 + 1 GR-like features (such as indefiniteness of the kinetic term).
ii) The Montesinos--Rovelli--Thiemann model mimics the algebra formed by the GR constraints more closely than the RPM does, 
but has no known whole-universe interpretation to make it appropriate as a cosmological toy model.  
The general idea of such models is that different ones closely resemble full GR in different ways, 
so that each is insightful under a different set of circumstances whilst remaining mathematically 
tractable unlike full GR, and building up concrete models of RPM's as done in this paper is part of this scheme.

\noindent 
In particular, RPM's have many further useful analogies \cite{K92, B94I, B94II, EOT, Paris, 06II, SemiclI, Records, 
08II, AF, Cones} with GR at the level of conceptual aspects of Quantum Cosmology, including the Problem 
of Time \cite{K81, K91, K92, I93, K99, Kieferbook, Smolin08}.
This notorious problem occurs because `time' is taken to have a different meaning in each of GR and 
ordinary quantum theory.  
This incompatibility creates serious problems with trying to replace these two branches of physics with 
a single framework in regimes in which neither QM nor GR can be neglected, such as in black holes or in 
the very early universe.  
One facet of the Problem of Time appears in attempting canonical quantization of GR due to ${\cal H}$ 
being quadratic but not linear in the momenta (a feature shared by RPM's energy constraint $H$). 
Then elevating ${\cal H}$ or $H$ to a quantum equation produces a stationary i.e timeless or frozen 
wave equation. 
In the GR case, this is the Wheeler-DeWitt equation 
\beq
\hat{\cal H}\Psi = -\left\{
\frac{1}{\sqrt{\cal M}}\frac{\delta }{\delta h_{ij}}
\left\{
\sqrt{\cal M}{\cal N}_{ijkl} 
\frac{\delta }{\delta h_{kl}}
\right\} 
+ \sqrt{h}R
\right\}\Psi = 0
\eeq 
(for $\Psi$ the wavefunction of the Universe). 
Note that one gets this frozen equation {\sl in place of} ordinary QM's time-dependent Schr\"{o}dinger 
equation, 
\beq
i\hbar\pa\Psi/\pa t = \hat{\fH}\Psi \mbox{ } .
\eeq 
[$t$ is here absolute Newtonian time; I use $\fH$ to denote Hamiltonians.] 
See \cite{K92, I93, APOT} for other facets of the Problem of Time.

Some of the strategies toward resolving the Problem of Time are as follows.  

\noindent A) 
Perhaps one is to find a hidden time at the classical level \cite{K92} by performing a canonical 
transformation under which the quadratic constraint is sent to $p_{t^{\th\ti\td\td\te\tn}} + 
\fH_{\st\sr\su\se} = 0$.  
Here, $p_{t^{\th\ti\td\td\te\tn}}$ is the momentum conjugate to some new coordinate 
$t^{\sh\si\sd\sd\se\sn}$ that is a candidate timefunction), which is then promoted to a 
$t^{\sh\si\sd\sd\se\sn}$-dependent Schr\"{o}dinger equation 
\beq
i\hbar\pa \Psi/\pa t^{\sh\si\sd\sd\se\sn} = \hat{\fH}_{\st\sr\su\se}\Psi \mbox{ } .
\eeq
York time \cite{York72, York73, K81, K92, I93} is a candidate internal time for GR, associated with 
constant mean curvature foliations. 

\noindent B)
Perhaps one has slow, heavy `$h$'  variables that provide an approximate timefunction with respect to 
which the other fast, light `$l$' degrees of freedom evolve \cite{HallHaw, K92, Kieferbook}.  
In Quantum Cosmology the role of $h$ is played by scale (and homogeneous matter modes), so scaled RPM's 
in scale--shape split are more faithful semiclassical models of this than pure-shape RPM's.   
In the Halliwell--Hawking set-up \cite{HallHaw}, the $l$-part are small inhomogeneities.  
(This is a more complicated model than the abovementioned ones, involving inhomogeneous 
perturbations about minisuperspace; its main drawback is that it is already very complicated 
as regards performing explicit Problem of Time strategy calculations; features 2) and 5) 
of RPM's make them useful toy models of \cite{HallHaw} in significant ways in which 
minisuperspace itself is not a good toy model.)
This approach goes via making the Born--Oppenheimer ansatz $\Psi(h, l) = 
\psi(h)|\chi(h, l)\rangle$ and the WKB ansatz $\psi(h) = \mbox{exp}(iW(h)/\hbar)$. 
One then forms the $h$-equation ($\langle\chi| \hat{H} \Psi = 0$ for RPM's), which, under a number 
of simplifications, yields a Hamilton--Jacobi equation\footnote{For simplicity, this is
presented in the case of one $h$ degree of freedom and with no linear constraints.} 
\beq
({\pa W}/{\pa \mh})^2 = 2(E - V(\mh)) \mbox{ } .  
\label{h-HJE}
\eeq
Here, $E$ is the total energy and $V(\mh)$ the $h$-part of the potential. 
Next, one approach to (\ref{h-HJE}) is to solve it for an approximate emergent semiclassical time 
$t^{\se\sm} = t^{\se\sm}(\mh)$. 
Then the $l$-equation $(1 - |\chi\rangle\langle\chi|)\hat{H}\Psi = 0$ can be recast (modulo further 
approximations) as a $t^{\se\sm}$-dependent Schr\"{o}dinger equation for the l degrees of freedom
\beq
i\hbar\pa|\chi\rangle/\pa t^{\se\sm}  = \widehat{H}_{l}|\chi\rangle \mbox{ }
\label{TDSE2}
\eeq
(where the left-hand side arises from the cross-term $\pa_{h}|\chi\rangle\pa_{h}\psi$ and 
$\widehat{H}_{l}$ is the remaining surviving piece of $\widehat{H}$ that serves as Hamiltonian 
for the $l$-subsystem).  

\noindent C) A number of approaches take timelessness at face value. 
One considers only questions about the universe `being', rather than `becoming', a certain way.  
This can cause some practical limitations, but can address at least some questions of interest. 
For example, the {\it na\"{\i}ve Schr\"{o}dinger interpretation} \cite{HP86,UW89} concerns the `being' 
probabilities for universe properties such as: what is the probability that the universe is large? 
Flat? 
Isotropic? 
Homogeneous?   
One obtains these via consideration of $\int_{R}|\Psi|^2\d\Omega$ for $R$ a suitable region of the 
configuration space and $\d\Omega$ is the corresponding volume element.  
This approach is termed `na\"{\i}ve' due to it not using any further features of the constraint 
equations.  
The {\it conditional probabilities interpretation} \cite{PW83} goes further by addressing conditioned 
questions of `being' such as `what is the probability that the universe is flat given that it is 
isotropic'?  
{\it Records theory} \cite{PW83, GMH, B94II, EOT, H99, Records} involves localized subconfigurations 
of a single instant.  
In particular, one is interested in whether these contain useable information, are correlated to each 
other, and whether a semblance of dynamics or history arises from this scheme.  
This requires notions of localization in space and in configuration space as well as notions of 
information.   
RPM's are superior to minisuperspace for such a study as they have i) a notion of localization in space. 
ii) They have more options for well-characterized localization in configuration space, i.e. of `distance 
between two shapes' \cite{NOD}.  
This is because RPM's have kinetic terms with positive-definite metrics, in contrast to GR's indefinite 
one.  

\noindent D) Perhaps instead it is the histories that are primary ({\it histories theory} \cite{GMH, 
Hartle}).    

\noindent E) For further completeness, I mention that distinct timeless approaches involve e.g. 1) 
\cite{Rovellibook} Rovelli's notion of {\it evolving constants of the motion} (which is underlied by a 
`Heisenberg' rather than `Schr\"{o}dinger' approach to QM). 
2) {\it Partial observables}.  
(The latter are used in Loop Quantum Gravity's {\it master constraint program} \cite{Thiemann}).  

However, my main interest is in combining B) to D) (for which RPM's are well-suited, due to 
possessing a nontrivial notion of locality). 
This is a particularly interesting prospect \cite{H03} for the following reasons (see \cite{ASharp} 
for more detailed arguments.) 
Firstly, there is a records theory within histories theory. 
Secondly, decoherence of histories is one possible way of obtaining a semiclassical regime in the first 
place. 
Thirdly, what the records are will answer the further elusive question of which degrees of freedom  
decohere which others in Quantum Cosmology.

\subsection{Outline of the rest of this paper}

Recent advances with scaled RPM stem from 

\noindent 
1) completion of the study of pure-shape RPM \cite{FORD, 08I, 
08II, AF, +tri, FileR} via the shape--scale split of scaled RPM \cite{08I, Cones}, by which pure-shape  
RPM is shown to occur as a subproblem within scaled RPM.  
Sec 2 provides coordinates and techniques for this, as well as the notation used in this paper.  
%

2) I present a Cosmology--Mechanics analogy in Sec 3.3 that serves to select what potentials the RPM's 
are to have if they are to more closely parallel known cosmological models.  
Under this analogy, mechanical energy corresponds to scalar curvature, harmonic oscillator (HO) terms 
(or their upside-down counterparts) to the cosmological constant, Newtonian gravity terms to dust and 
conformal potential terms to radiation.
I also cover how the `scale-dominates-shape' approximation (c.f. Sec \ref{SSA}), that is linked 
to the RPM version of the Cosmology--Mechanics analogy, is not always stable.  
I find that some of the potentials arising from this analogy have the further useful features of being 
more analytically tractable and well-behaved at the quantum level.   
I view these useful properties as acting as further filters for selecting appropriate models for 
further QM and toy-modelling of Problem of Time strategies.  
By considering scale and working through to the quantum level, the present paper is a natural 
sequel of both \cite{Cones} and \cite{AF}.  
It also underlies an upgrade of the semiclassical approach in \cite{SemiclI} to the Problem of Time and 
Quantum Cosmology, by having scale as its $h$-part part that provides an approximate emergent 
timefunction with respect to which the $l$-dynamics of pure shape runs.  
Along these lines, RPM's can be used as a toy model of the Halliwell--Hawking set-up \cite{HallHaw}.

This paper's approach to quantization is laid out in Sec 4.
Secs 5--7 then find and study exact solutions to the free and multi-HO potential cases, as well as 
approximate solutions under the scale-dominates-shape approximation for a wider range choice of potentials 
inspired by the Cosmology--Mechanics analogy.  
These are treated for 3- and 4-stop metroland in the main text, and are extended to $N$-stop metroland in 
\cite{v2}.
[I call the RPM of $N$ particles on a line $N$-stop metroland, because their configurations 
look like the familiar depiction of public transport lines. 
$N$-a-gonland is the RPM of $N$ particles in 2-$d$, the 3- and 4- particle cases of which 
I refer to as triangleland and quadrilateralland.] 
These problems give familiar Bessel, Hermite and Laguerre function    mathematics, albeit this now admits a new 
interpretation in terms of ratios of relative particle separations, as opposed to particle positions. 
This makes it appropriate for whole-universe modelling.  
Also note that scalefree RPM calculations in \cite{AF, FileR} are relevant again through reappearing, 
via the shape--scale split, as the shape parts of the present paper's scaled RPM calculations. 
Some of these solutions admit a characteristic (mass-weighted) length, i.e. (the square root of) a 
`Bohr moment of inertia' and an HO lengthscale. 
I also interpret these solutions by use of expectations and spreads (paralleling \cite{AF, +tri}), 
and point out that a wider range of cases can be studied using fairly standard perturbation theory.

In the conclusion (Sec 8), I discuss applications, particularly to the semiclassical approach to 
the Problem of Time but also to hidden time and timeless approaches (the na\"{\i}ve Schr\"{o}dinger 
interpretation and records theory) and to robustness of, and notions of uniformity in, Quantum Cosmology.

\section{Classical set-up for scaled $N$-stop metroland models}

\subsection{Useful coordinatizations of the configuration spaces}

Consider $N$ particles in dimension $d$, with position coordinates ${\bf q}^a$ and masses $m_a$.
A configuration space for these is $\fQ(N, d) = \mathbb{R}^{Nd}$.  
However, this includes physically-meaningless information about absolute position and the 
absolute orientation of the axes.  
Rendering absolute position irrelevant (by passing to any sort of relative coordinates) 
straightforwardly leaves one on a configuration space {\it relative space} = $\fR(n, \mbox{ } d) = 
\mathbb{R}^{nd}$.
The most obvious relative space coordinates formed from these are some basis set among the 
${\bf r}^{ab} = {\bf q}^b - {\bf q}^a$, however certain linear combinations of these 
-- {\it relative Jacobi coordinates} \cite{Marchal} prove to be more advantageous to use.  
I denote these by ${\bf R}^e$.
These are physically relative separations between {\sl clusters} of particles (see Fig 1 for the 
particular examples of these used in this paper). 
The main advantage of using the ${\bf R}^e$ is that the kinetic metric is then diagonal.  
Furthermore, it looks just like (see Sec 3.1) the kinetic metric in terms of the ${\bf q}^a$'s but involving one 
object less and with cluster masses $\mu_e$ in place of particle masses $m_a$.
In fact, in this paper I use 1) {\it mass-weighted} relative Jacobi coordinates $\brho^e = 
\sqrt{\mu_e}{\bf R}^e$. 
2) Their squares the partial moments of inertia $I^e = \mu_e|{\bf R}^e|^2$. 
3) `{\it normalized}' versions of the former, ${\bf n}^e = \brho^e/\rho$, 
where $\rho := \sqrt{I}$ is the {\it hyperradius}, $I$ itself being the {\it total moment of inertia} 
of the system. 
For specific coordinate components, I write the index downstairs since this often turns out to 
simplify the presentation.

{            \begin{figure}[ht]
\centering
\includegraphics[width=0.75\textwidth]{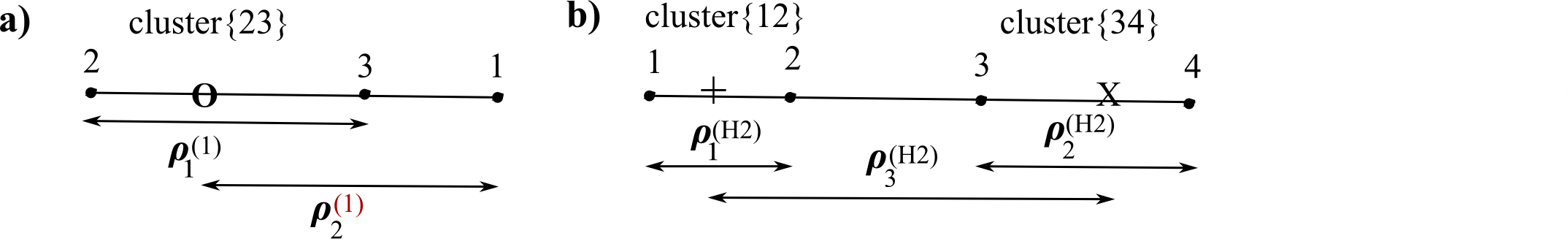}
\caption[Text der im Bilderverzeichnis auftaucht]{        \footnotesize{O, +, and X are the centres of 
mass of particles 2 and 3, 1 and 2, and 3 and 4 respectively.

\noindent a) For 3-stop metroland, there are 3 permutations that correspond to following 
each of the particle clusterings \{1, 23\}, and cycles.  
I draw the first of these, which I denote by the shorthand $(1)$.  

\noindent b) One particular sort of relative Jacobi coordinate system for 4-stop metroland that 
correspond to following the clustering \{1b, cd\} `into pairs 1b and cd' where b, c, d form a cycle. 
I term these H-clusterings since in dimension $\geq 2$ they form the shape of the letter H.  
I draw the first of these, which I denote by the shorthand (H2) as particle 1 is paired with particle 2. 
The other sort of relative Jacobi coordinate system for 4 particles are K-coordinates, corresponding 
to following clustering \{a, bcd\} (see \cite{FileR, QSub} for more on these).  
That each coordinate system just `follows one clustering', means that in general we need to use various 
coordinate systems (this also makes sense from a differential-geometric perspective).  
The present paper just works in (1) and (H2) coordinates, so I subsequently drop clustering labels.  
}        }
\label{Fig1}
\end{figure}            }

Eliminating rotations is harder.  
Doing so sends one to {\it relational space} ${\cal R}(N, d)$.  
Eliminating scale as well sends one to {\it shape space}, $\fS(N, d)$. 
Eliminating scale {\sl instead} of rotations sensd one to {\it preshape space}, $\fP(N, d)$.
One can then consider relational space as the {\it cone} $\mC(\fS(N, d))$ over shape space; this 
amounts to applying a shape--scale split.  
In the 1-$d$ models of the present paper, if we take the ordering abc to be distinct from cba (or abcd 
to be distinct form dcba), our shapes are `plain' rather than `orientation-identified'.  
As the current paper is in 1-$d$, there are no rotations, so ${\cal R}(N, d) = \fR(N, d)$. 
This case is simpler, but makes for a useful precursor and already has a number of Problem of Time and 
quantum cosmological applications as presented in the Conclusion.
Taking relative space to correspond to the space of Riemannian 3-metrics on a fixed spatial topology $\Sigma$  
(taken to be compact without boundary for simplicity), then relational space corresponds to Wheeler's superspace($\Sigma$) 
\cite{Wheeler}, shape space to conformal superspace CS($\Sigma$) and preshape space to pointwise conformal superspace \cite{FM96}.
There is also an analogy between the cone over shape space and 

\noindent\{CS + V\}($\Sigma$), where the V stands for a single global 
3-volume degree of freedom \cite{York73, ABFKO}.

In the present paper's 1-d context, 
${\cal R}(N, d) = \mathbb{R}^{n}$, $\fS(N, d) = \mathbb{S}^{n - 1}$, and the shape-scale 
split amounts to using the $\mathbb{S}^{n - 1}$ analogue of spherical polar coordinates.  
E.g. for scaled 3-stop metroland I use $\{\rho, \varphi\}$ coordinates with 
\beq
\rho = \sqrt{\rho_1^2 + \rho_2^2} \mbox{ } , \mbox{ } \mbox{ } 
\varphi = \mbox{arctan}(\rho_1/\rho_2) \mbox{ } , 
\eeq
while for scaled 4-stop metroland I use \{$\rho, \theta, \phi$\} coordinates with 
\beq
\rho = \sqrt{\rho_1^2 + \rho_2^2 + \rho_3^2} 
\mbox{ } , \mbox{ } \mbox{ } \theta = \mbox{arctan}(\sqrt{\rho_1^2 + \rho_2^2}/\rho_3)
\mbox{ } , \mbox{ } \mbox{ } \phi   = \mbox{arctan}(\rho_2/\rho_1) \mbox{ } .  
\eeq
For 3-stop metroland \cite{AF}, $n_1^{(\sH \sb)}$ is a measure of how large cluster bc is relative to 
the size of the model universe, and $n_2^{(\sH \sb)}$ is a measure of how large the separation 
between cluster bc and particle a is relative to the size of the model universe. 
For 4-stop metroland \cite{AF}, $n_1^{(\sH \sb)}$ is a measure of how large cluster 1b is 
relative to the size of the model universe.   
$n_2^{(\sH \sb)}$ is a measure of how large cluster cd is relative to the size of the model universe. 
Finally, $n_3^{(\sH \sb)}$ is a measure of how large the separation between clusters 
1b and cd is relative to the size of the model universe.  
$\rho$ itself is a scale variable.    
In this paper I restrict attention to the case of equal masses for simplicity and for homogeneity of the 
contents of the model universe.
%

\subsection{Useful tessellations of the configuration spaces}

\noindent Tessellations of the configuration space are useful for reading off the physical 
interpretation, i.e. what shape the particles make. 
Thus, one can readily interpret classical trajectories as paths on the tessellated configuration space, 
and classical potentials and quantum-mechanical probability density functions as height functions over 
it.

For plain scaled 3-stop metroland \cite{06I, Cones}, in the plain case (the rim of Fig 2a), the shape 
space is the circle with 6 regularly-spaced double collision (D) points on it.  
It also has 6 somewhat less significant points -- {\it mergers}, M: where the centre of mass 
of 2 particles coincides with the third particle.  
These merger points lie halfway between each two adjacent D's so that the D and M points form an 
hour-marked clock face.  
The 6 D's pick out 3 preferred axes, and the 6 M's pick out 3 further preferred axes perpendicular 
to the first 3.    
Each pair of D's and perpendicular pair of M's corresponds to one of the 3 permutations of Jacobi 
coordinates.   
The polar angle $\varphi^{(\sa)}$ about each of these axes is then natural for the study of the 
clustering corresponding to that choice of Jacobi coordinates, \{a, bc\} = (a).  
The relational space is then the cone over this decorated shape space, i.e. (the infinite extension of) 
Fig 2a)'s pie of 12 slices with 6 D half-lines and 6 M half-lines.   
All of these emanate from the triple collision at the cone point, 0.  
The tesselation method also works for triangleland \cite{+tri} (albeit the tesselation in question 
is very different), and is also important in the study of quadrilateralland \cite{QSub}.

{            \begin{figure}[ht]
\centering
\includegraphics[width=0.5\textwidth]{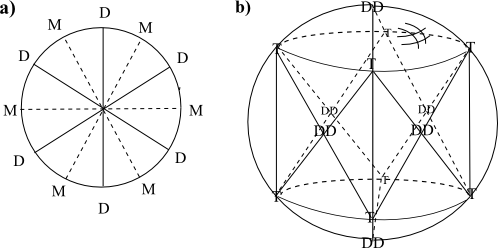}
\caption[Text der im Bilderverzeichnis auftaucht]{        \footnotesize{A sketch of a) the tessellation 
of 3-stop metroland's relational space (the corresponding shape space is the rim of this). 
b) Of 4-stop metroland's shape space (the corresponding relational space is the cone over this).}     }
\label{Fig2}
\end{figure}  }

For plain 4-stop metroland \cite{AF, +tri} (Fig 2a), the shape sphere is tessellated by 24 isosceles 
spherical triangle faces with 14 vertices. 
8 of these are triple collision (T) points and 6 are  are double-double (DD) collision points. 
There are 36 edges, which are lines of double collisions (D). 
Each DD is attached to 4 T's, and each T to 3 T's and 3 DD's, so that the T's and D's form, 
respectively, the vertices of a cube and of the octahaedron dual to it.     
The vertices also form 7 antipodal pairs, thus picking out 7 preferred axes (3 pairs of DD's  
corresponding to the permutations of Jacobi H-coordinates and 4 pairs of T's corresponding to 
the permutations of Jacobi K-coordinates.
This correspondence is in the sense that the poles in each case are where what each coordinatization 
picks out as intra-cluster coordinates both go to zero -- collapse of both clusters for an H-clustering 
or collapse of the triple cluster for a K-clustering.  
Then the spherical polar coordinates about each principal axis -- a \big($\phi^{(\sH\sb)}$, 
$\theta^{(\sH\sb)}$\big) or a \big($\phi^{(\sK\sa)}$, $\theta^{(\sK\sa)}$\big) coordinate system -- are 
natural for the study of the clustering corresponding to that choice of Jacobi coordinates.  
I.e. each choice of H- or K-clustering has a different natural spherical polar coordinate chart,   
any two of which suffice to form an atlas for the sphere.   
%
%
The relationalspace for the scaled theory is then the solid cone made from each corresponding shape 
space decorated by its tessellation.  
This has, furthermore, a maximal (quadruple) collision at the cone point, 0.

\section{The classical problem}

\subsection{Relational actions}

The relational Jacobi-type action \cite{Lanczos} for scaled RPM is
\beq
S = 2\int\sqrt{T(E - V)}\d\lambda \mbox{ } 
\eeq
where the kinetic term $T$ is, in the particle position presentation, 
\beq
T = \sum\mbox{}_{\mbox{}_{\mbox{\sI}}}
\frac{    m_a\{\mq^{ai\prime}    - \ma^{i\prime} - \{\mb^{\prime}\cr\mq^{a}\}^{i}\}
             \{\mq^{a\prime}_{i} - \ma^{\prime}_i - \{\mb^{\prime}\cr\mq^{a}\}_{i}\}     }{2} = 
\frac{    m_a\delta_{ab}\delta_{ij}
             \{\mq^{ai\prime} - \ma^{i\prime} - \{\mb^{\prime}\cr{\mq}^a\}^{i} \}
             \{\mq^{bj\prime} - \ma^{j\prime} - \{\mb^{\prime}\cr{\mq}^b\}^{j}  \}   }{2} 
                                                        \mbox{ } 
\eeq
(using the Einstein summation convention in the second expression, and where $\ma^{\alpha}$ and 
$\mb^{\alpha}$ are translational and rotational auxiliary variables, and where $\lambda$ is a label and 
$^{\prime} := \d/\d\lambda$).    
In the relative Jacobi coordinates presentation, the kinetic term is 
\beq
T = \sum\mbox{}_{\mbox{}_{\mbox{\si}}} 
\frac{    \mu_e\{{\mR}^{ei\prime} - \{{\mb}^{\prime}\cr{\mR}^e\}^{i}\}
               \{{\mR}^{e\prime}_i - \{{\mb}^{\prime}\cr{\mR}^e\}_{i}\}    }{2} = 
\frac{    \mu_e\delta_{ef}\delta_{ij}
               \{{\mR}^{ei\prime} - \{{\mb}^{\prime} \cr {\mR}^e\}^{i}\}
               \{{\mR}^{fj\prime} - \{{\mb}^{\prime} \cr {\mR}^f\}^{j}\}   }{2} 
                                                        \mbox{ } .
\eeq
In the reduced case,\footnote{${\cal R}^{q}$ are relational space coordinates with
conjugate momenta ${\cal P}_{q}$. 
${\cal M}_{pq}$ is the relational space metric with determinant ${\cal M}$ and inverse ${\cal N}^{pq}$.  
$S^{u}$ are shape space coordinates with conjugate momenta $P_{u}$. 
$M_{uv}$ is the shape space metric with determinant ${M}$ and inverse ${N}^{uv}$.}  
\beq
T = {\cal M}_{pq}{\cal R}^{p\prime}{\cal R}^{q\prime}/2 = T_{\rho} + 
\rho^2\fT_{S} \mbox{ } 
\eeq
where 
\beq
T_{\rho} = \rho^{\prime\,2}/2 \mbox{ } , \mbox{ } \mbox{ } 
\fT_{S} = M_{uv}{S}^{u\prime}{S}^{v\prime} = 
\sum\mbox{}_{\mbox{}_{\mbox{\scriptsize $r = 1$}}}^{n - 1}
\prod\mbox{}_{\mbox{}_{\mbox{\scriptsize $s = 1$}}}^{r - 1}
\mbox{sin}^2\theta_{s}{\theta}_{r}^{\prime\,2}  /2 
\eeq
are scale and shape parts respectively, and where the last equality is for the specific case of 
$N$-stop metroland in terms of ultraspherical angles \{$\theta_{r}$, $r$ = 1 to $N$ -- 2\}).
In particular, for 3-stop metroland, 
\beq
\fT_{S} = {\varphi}^{\prime\,2}/2
\eeq 
and for 4-stop metroland,  
\beq
\fT_{S} = ({\theta}^{\prime\,2} + \mbox{sin}^2\theta\,{\phi}^{\prime\,2})/2 \mbox{ } . 
\eeq  
These actions implement temporal relationalism via reparametrization invariance, since $\fT$ is purely quadratic in 
$d/d\lambda$ and occurs as a square root factor so that these $\d/\d\lambda$'s cancel with the $\d\lambda$ of the 
integration and thus the $\lambda$ is indeed a mere label.
These actions implement configurational relationalism via the corrections to the $\mq^{a\prime}$ and $\mR^{e\prime}$ 
(the linear constraint coming from variation with respect to $\mb^{\alpha}$ is \ref{ZAM}) and via 
using G-invariant constructs directly in the reduced approach.

The GR counterpart, for comparison and to further substantiate the tightness of the GR--RPM analogy is an also reparametrization-invariant 
and thus temporal relationalism implementing action that is a variant of the Baierlein--Sharp--Wheeler action \cite{BSW,RWR},   
\beq
S = \int\d\lambda\int\sqrt{h}\sqrt{T_{\sG\sR}\{- 2\Lambda + R\}}
\eeq
where
\beq
T_{\sG\sR} = M^{ijkl}\{{h}^{\prime}_{ij}     - \pounds_{\sF^{\prime}}h_{ij}    \}
                                   \{{h}^{\prime}_{kl} - \pounds_{\sF^{\prime}}h_{kl}\} \mbox{ } .  
\eeq
Here, $M^{ijkl}$ is the undensitized version of the GR configuration space metric (of which the 
DeWitt supermetric is the inverse), equal to $h^{ik}h^{jl} - h^{ij}h^{kl}$,
$\sF^{\prime}$ plays the same role mathematically as the GR shift and 
$\pounds_{\sF^{\prime}}$ is the the Lie derivative with respect to $\mF^{\prime}$.
The $\pounds_{\sF^{\prime}}h_{ij}$ corrections to the metric velocities then indirectly implement 
configurational relationalism with respect to the 3-diffeomorphisms; the associated constraint is (\ref{GRmom}).

The momentum--velocity relations and the equations of motion for the RPM's then feature the combination

\noindent $\sqrt{(E - V)/T}\pa/\pa\lambda := \d/\d t^{\se\sm} := \dot{\mbox{ }}$, for 
$t^{\se\sm}$ the {\it emergent time} of the relational approach.  
Such an emergent time amounts to a recovery of Newtonian, proper and cosmic time in various different 
contexts, as well as coinciding with the semiclassical notion of emergent time mentioned around eq (5).  
I then denote this joint notion of emergent time by $t^{\se\sm}$. 
See \cite{Cones} for the equations of motion.

\subsection{Conserved quantities and energy equation for $N$-stop metroland}

If $V$ is shape-independent, there are conserved quantities. 
For $N$-stop metroland, one gets an SO($N$ -- 1)'s worth of these.   
Denote these by ${\cal D}_{\Delta}$.  
For $N$ = 3 there is one, ${\cal D} = \rho^2\dot{\varphi}$.
For $N$ = 4, there are three ${\cal D}_{\Delta}$, which can be taken to have a relative quantity 
or relational space index, ${\cal D}_{e}$. 
Only one of these is conserved if the potential were to depend on $\theta$ but not $\phi$.  
Note that, while the ${\cal D}_{\Delta}$ have $(N - 1)$-$d$ angular momentum {\sl mathematics},  
they are not {\sl physically} angular momenta.    
More specifically, in $N$-stop metroland, these conserved quantities are 
{\it relative dilational momenta} \cite{AF}.
(E.g. $R_1P_1$ is the dilational momentum of particle cluster 12; for $d > 1$, 
dilational momenta are dot, as opposed to cross, products.)
The $N$ = 3 model has a tight mathematical analogy with the central force problem in the plane whose 
sole conserved quantity is the angular momentum, $L$. 
The $N$ = 4 problem is an analogy with the central force problem in spherical coordinates, in which case 
the ordinary angular momenta $L_i$ play the role of the ${\cal D}_{i}$.  
In 2-$d$, the conserved quantities that arise are physically a mixture of relative dilational momentum 
and relative angular momentum \cite{+tri}. 
The physical generalization of angular momenta \cite{Smith} that comprises all of angular momenta, 
dilational momenta, and mixtures of these, can be interpreted as {\it rational momenta} 
(i.e., associated with ratios, of which angles are but one example) \cite{AF, +tri, Cones}
In  each case, there is a ${\cal D}_{\sT\so\st}$; for $N$ = 3 ${\cal D}_{\sT\so\st} = {\cal D}^2$, while 
for $N$ = 4 ${\cal D}_{\sT\so\st} = \sum_{i = 1}^3{\cal D}_{i}\mbox{}^2$.

The general problem has an energy relation 
\beq
E - V(\rho, S^{u}) = (\dot{\rho}^{2} + 
\rho^2M_{uv}\dot{S}^{u}\dot{S}^{v})/2  =  
(P_{\rho}^2 + N^{uv}P_{u}P_{v}/\rho^2)/2 = {\cal N}^{pq}{\cal P}_{p}{\cal P}_{q}/2 
\mbox{ } 
\label{EnRel}
\eeq
for $P_{\rho}$ the momentum conjugate to the hyperradius scale variable, $\rho$.    
Using the conserved quantities, we can write this as   
\beq
\dot{\rho}^{2}/2 + {\cal D}_{\sT\so\st}/2\rho^2 + V(\rho) = E \mbox{ } .
\eeq

\subsection{What potentials to use in RPM's: Cosmology--RPM analogy}\label{SSA}

Pure-shape RPM has a lot of potential freedom (any potential homogeneous of degree 0), while scaled RPM 
can have any potential at all.
Hitherto, the RPM program has covered free problems and HO potentials (or HO-like ones for 
pure-shape RPM, for which HO potentials themselves are disallowed by the above homogeneity requirement).   
In the scaled case, these are $\rho^2(A + B\,\mbox{cos}\,2\varphi)$ for 3-stop metroland and 
$\rho^2(A + B\,\mbox{cos}\,2\theta + C\,\mbox{sin}^2\theta\,\mbox{cos}\,2\phi)$ for 4-stop metroland.  
The advantages of such potentials are boundedness and good analytical tractability.  
However, such models are atypical in their simpleness and do not particularly parallel the 
dominant scale dynamics of commonly-used cosmological models.

The energy relation (\ref{EnRel}) also holds approximately if there is shape dependence, so long as it 
and changes in it are small compared to changes of scale. 
This requires some stability condition so as to apply long-term. 
In this case one has $V_{(0)}$ in place of $V$.   
Obtaining a separated-out heavy slow scale part for the semiclassical approach relies on the following 
`scale-dominates-shape' approximation being meaningful,
\beq
|T_{l}| << |T_{h}| \mbox{ realized by the shape quantity } \rho^2\fT_{S} <<   T_{\rho} 
\mbox{  (scale quantity) } ,
\label{SSA1}
\eeq
\beq
|J_{hl}| << |V_{h}| \mbox{ realized by } |J(\rho, S^{u})| << |V_{(0)}(\rho)|  
\mbox{ for }
\label{SSA2}
\eeq
for $J_{hl}$ the interaction part of the potential.  
\beq
V(\rho, S^{u}) \approx V_{(0)}(\rho)(1 + V_{(1)u}(\rho)S^{u} + 
O(|S^{u}|^2) = V_0(\rho) + J(\rho, S^{u})  ) \mbox{ } .  
\label{SSA3}
\eeq 
[Without such approximations, one cannot separate out the heavy (here scale) part so that it 
can provide the approximate timefunction with respect to which the light (here shape) part's 
dynamics runs.]
Counterparts of this are e.g. the leading-order neglect of scalar field terms in e.g. \cite{GibGri}, of anisotopy 
in e.g. \cite{Amsterdamski} and of inhomogeneity in e.g. \cite{HallHaw}.  

\mbox{ } 

As previously highlighted, a main purpose of this paper is to use the analogy between Mechanics and 
Cosmology to broaden the range of potentials under consideration and pinpoint ones which parallel 
classical and Quantum Cosmology well. 
This makes more profitable use of the potential freedom than previous mere use of simplicity.

Isotropic cosmology (in $c = 1$ units) has the Friedmann equation
\beq
\left(\frac{\dot{a}}{a}\right)^2 = - \frac{k}{a^2} + \frac{8\pi G\varepsilon}{3} + \frac{\Lambda}{3}  
= -\frac{k}{a^2} + \frac{2GM_{\sd\su\sss\st}}{a^3} + \frac{2GM_{\sr\sa\sd}}{a^4} + \frac{\Lambda}{3}
\mbox{ } ,    
\eeq
the second equality coming after use of energy--momentum conservation and assuming noninteracting matter 
components. 
Here, $a$ is the scalefactor of the universe, $\dot{\mbox{ }} = \d/\d t^{\sc\so\sss\sm\si\sc} ( = \d/\d t^{\se\sm}$ 
here). 
$k$ is the spatial curvature which is without loss of generality normalizable to 1, 0 or --1. 
$G$ is the gravitational constant, $\varepsilon$ is matter energy density, $\Lambda$ is the cosmological 
constant and $M$ is the mass of that matter type enclosed up to radius $a(t)$.

A fairly common analogy is then between this and (unit-mass ordinary mechanics energy equation)$/r^2$, 
\beq
\left(\frac{\dot{r}}{r}\right)^2 = \frac{2E}{r^2} + \frac{K_{\mbox{\scriptsize Newton}}}{r^3} + 
\frac{K_{\mbox{\scriptsize conformal}}}{r^4} + \frac{K_{\mbox{\scriptsize Hooke}}}{3}
\mbox{ }    
\eeq
where here and elsewhere in this paper the various $K$'s are constant coefficients, and 
$\dot{\mbox{ }} = \d/\d t^{\se\sm}$ is now also $\d/d t^{\sN\se\sw\st\so\sn}$.  
A particularly well-known subcase of this is that 1-$d$ mechanics with a $1/r$ Newtonian gravity type 
potential is analogous to isotropic GR cosmology of dust. 
This extends to an analogy between the Newtonian dynamics of a large dust cloud and the GR isotropic 
dust cosmology \cite{CosMech,BGS03}.
(Here, shape is least approximately negligible through its overall averaging 
out to approximately separated out shape and cosmology-like scale problems.)  
For \cite{Cones} and the present paper's purposes, enough of these parallels (\cite{HarMTWPee, Rindler}) 
survive the introduction of a pressure term in the cosmological part, and the use of $N$-stop 
metroland RPM in place of ordinary mechanics.

The $N$-stop metroland--Cosmology analogy\footnote{This also applies to the $\mathbb{CP}^{N - 2}$ 
presentation of $N$-a-gonland.}
is between the above Friedmann equation and (\ref{EnRel})/$\rho^2$,
\beq
\left(\frac{\dot{\rho}}{\rho}\right)^2 =  
                                             \frac{2E}{\rho^2} 
                                             - \frac{{\cal D}_{\sT\so\st}}{\rho^4} 
                                             - \frac{2V(\rho, S^{u})}{\rho^2} = 
\frac{2E}{\rho^2} + \frac{2K}{\rho^3} + \frac{2R - {\cal D}_{\sT\so\st}}{\rho^4} - 2A \mbox{ } . 
\eeq 
Thus the spatial curvature term $k$ becomes --2 times the energy $E$.   
The cosmological constant term $\Lambda/3$ becomes --2 times the net $A$ from the (upside down) HO 
potentials.
The dust term $2GM/a^3$'s coefficient $2GM$ becomes --2 times the net coefficient $K$ from the 
Newtonian gravity potential terms. 
N.B. the $K$ of interest has the same sign for dust and for Newtonian gravity (and for the attractive 
subcase of Coulomb).   
The coefficient of the radiation term $2GM/a^4$, i.e. $2GM$, becomes $- {\cal D}_{\sT\so\st} + 2R$, where  
$2R$ the coefficient of the $V_{(0)}$ contribution from the $1/|{r}^{ab}|\mbox{}^2$ terms.  
Note that ${\cal D}_{\sT\so\st}$ itself is of the wrong sign to match up with the ordinary radiation 
term of cosmology.  
In the GR cosmology context, `wrong sign' radiation fluid means that it still has $p = \varepsilon/3$ 
equation of state (for $p$ the pressure), but its energy density $\varepsilon$ is negative.  
Thus it violates all energy conditions, making it unphysical in a straightforward GR cosmology context, 
and also having the effect of singularity theorem evasion by `bouncing'.  
From the ordinary mechanics perspective, however, this is just the well-known repulsion of the 
centrifugal barrier preventing collapse to zero size.
This difference in sign is underlied by an important limitation in the Mechanics--Cosmology analogy: in 
mechanics the kinetic energy is positive-definite, while in GR the kinetic energy is indefinite, with 
the scale part contributing negatively.  
This does not affect most of the analogy because the energy and potential coefficients can be considered  
to come with the opposite sign.  
However, the relative sign of the shape and scale kinetic terms cannot be changed thus, and it is from 
this that what is cosmologically the `wrong sign' arises. 
Two distinct attitudes to wrong-sign term are as follows.  
Firstly, one can suppress this ${\cal D}_{\sT\so\st}$ term by taking toy models in which this term 
(whose significance is the relative dilational momentum of two constituent subsystems) is zero or small. 
Alternatively, one can consider cases in which it is swamped by `right-sign' $1/|{r}^{ab}|\mbox{}^2$ 
contributions to the potential. 
Secondly, it should also be said that more exotic geometrically complicated scenarios such as brane 
cosmology can possess `dark radiation' including of the `wrong sign' \cite{DarkRad}. 
(Indeed, due to projections of higher-dimensional objects \cite{AT05}, it can possess what appears to be 
energy condition violation from the 4-$d$ perspective).  
Thus such a term is not necessarily unphysical.

What the Cosmology--RPM mechanics analogy provides at the semiclassical level is a slow heavy scale 
dynamics paralleling that of Cosmology.  
However, it is now coupled to a light fast shape dynamics that is simpler 
than GR's and has the added advantage of a meaningful notion of locality/inhomogeneity/structure.
The analogy does not however have a metric interpretation or a meaningful interpretation in terms of an 
energy density $\varepsilon$, it is after all just a particle mechanics model.

\subsection{Types of behaviour of (approximately) classical solutions}

{            \begin{figure}[ht]
\centering
\includegraphics[width=0.9\textwidth]{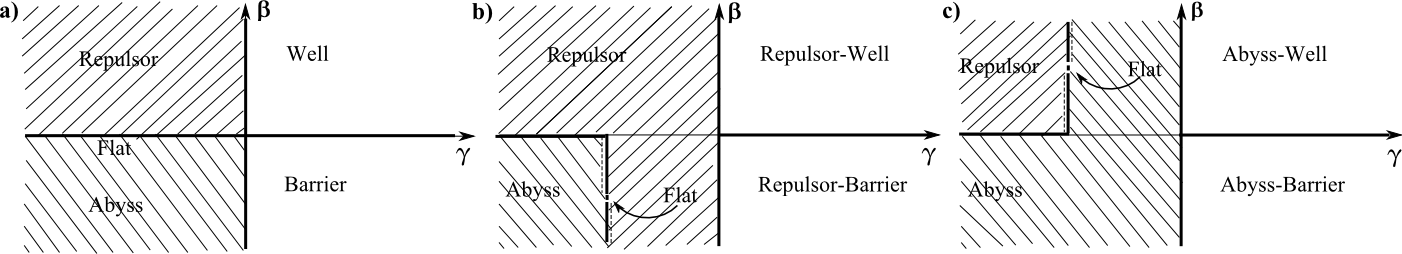}
\caption[Text der im Bilderverzeichnis auftaucht]{        \footnotesize{a) 5 classical behaviours for 
$V = \beta\rho^{\gamma} $ potentials.  
Solid lines denote included edges and dashed lines denote excluded edges.
In the top-right quadrant, our representative is the HO.  
In the bottom-right quadrant, our representative is the upside-down HO.  
In the bottom-left quadrant, our representative `abyss' is the bound l = 0 analogue of the hydrogen atom 
model.
In the top-left quadrant, our representative `repulsor' is the l = 0 analogue of the electron--electron model.
The fifth region is the axes, for which the potential is constant.  

\noindent b) 
Next consider the also commonly occurring (under the $\rho \longrightarrow r$, ${\cal D}_{\tN\te\ttt} = 
{\cal D}_{\tT\to\ttt} - 2R \longrightarrow L_{\tT\to\ttt\ta\tl}$  analogy) case of $V = \beta
\rho^{\gamma} + 
{\cal D}_{\tN\te\ttt}/\rho^2$, i.e. with a ${\cal D}_{\tN\te\ttt} > 0$ `centrifugal barrier' added.
This sends the HO to the repulsor HO, the upside-down HO to the repulsor upside-down HO. 
Also now the repulsor behaviour takes over the axes and a strip plus partial boundary of the 
bottom-left quadrant, pushing out the `abyss' behaviour to $\gamma < -2$ and the lower part of $\gamma 
= 2$.  
The only case exhibiting the flat behaviour now is the critical value of $\beta$ for $\gamma = - 2$.  

\noindent Note that one of our repulsor representatives, however, (l $\neq$ 0 bound Hydrogen analogue) 
has a well next to the repulsor.

\noindent c) Finally, consider the ${\cal D}_{\tN\te\ttt} < 0$ case c), which is not mechanically 
standard but is cosmologically standard (radiation term).  
This sends the HO to the abyss--HO, the upside-down HO to the abyss--upside-down-HO.  
Also now it is the abyss behaviour that takes over the axes and a strip plus partial boundary of the 
top-left quadrant.  
This pushes out the `repulsor' behaviour to $\gamma < -2$ and the upper part of $\gamma  = 2$.
}        }
\label{Fig3}
\end{figure}  }

\noindent 
It is important to have a grasp of the classical behaviour of a system prior to studying it  
quantum-mechanically. 
The shape of the potentials can be classified into the regions of Fig 3. 
The corresponding qualitative classical behaviours of 

\noindent these are in Fig 4; therein I make use of a notation along the lines of that of Robertson 
\cite{Robertson33} (also used in e.g. \cite{Harrison, Rindler}) for types of solution.    
This involves using 
$O$ for oscillatory models of the types 
$O_1$ ($0 \leq \rho \leq \rho_{\sm\sa\sx}$), 
$O_2$ ($\rho_{\sm\si\sn} \leq \rho \leq \rho_{\sm\sa\sx}$) and  
$O_3$ ($\rho_{\sm\si\sn} \leq \rho \leq \infty$).   
$M$ for monotonic solutions of type 
$M_1$ ($0 \leq \rho \leq \infty$), and 
$M_2$ ($\rho_{\sm\si\sn} \leq \rho \leq \infty$).   
$S$ for static solutions of types 
$S_1$ (unstable), 
$S_2$ (stable) and 
$S_3$ (stable for all $\rho$). 
$A$ for solutions asymptotic to static solutions at a finite $\rho_{\sA}$, of types 
$A_1$ coming in from 0 and 
$A_2$ coming in from $\infty$.

{            \begin{figure}[ht]
\centering
\includegraphics[width=0.8\textwidth]{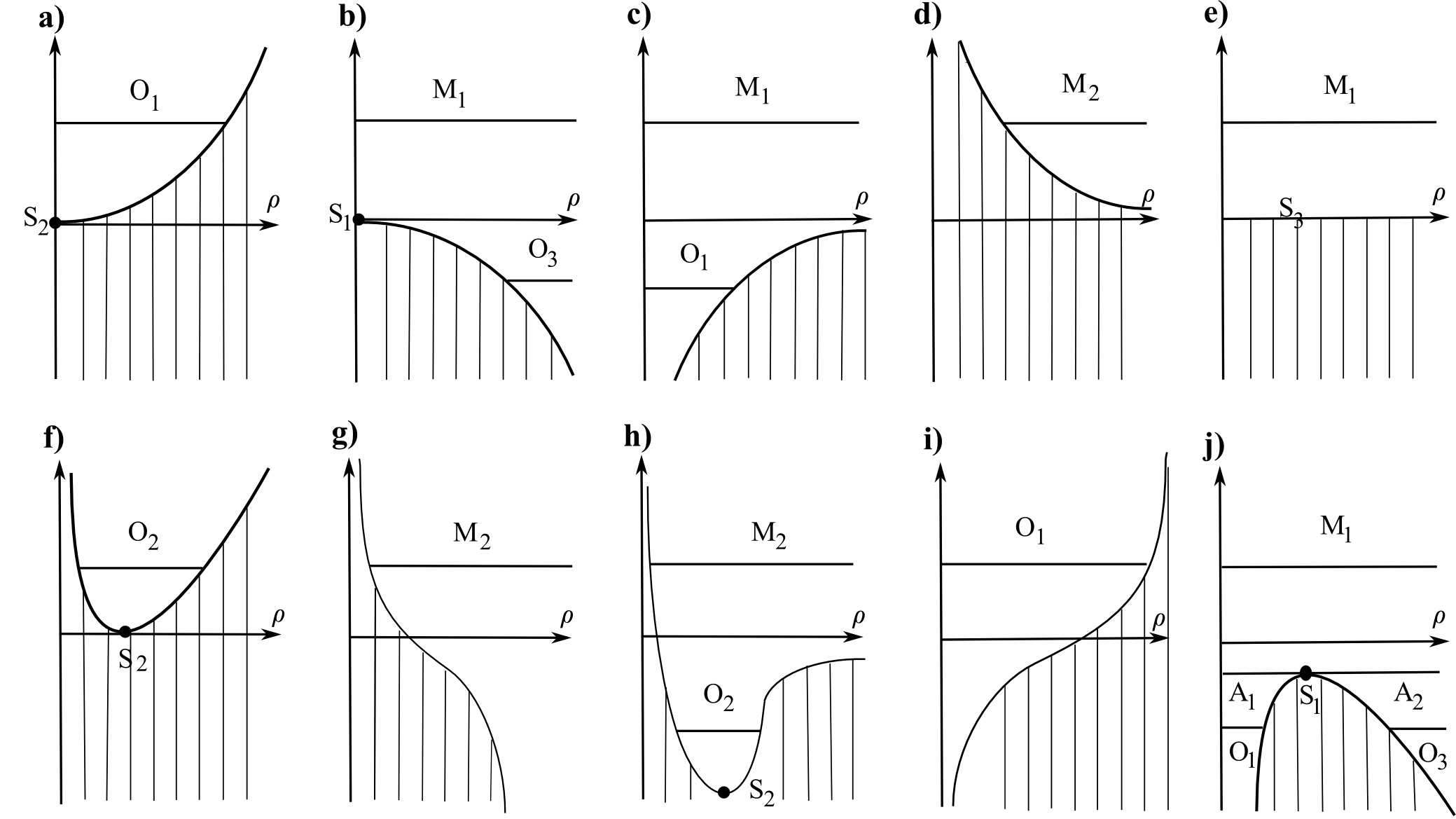}
\caption[Text der im Bilderverzeichnis auftaucht]{        \footnotesize{ Qualitative classical 
behaviour and identification of corresponding cosmological solutions.
Shaded regions are classically forbidden.

\noindent a) The HO model corresponds to a Milne in AdS cosmology; the potential is nonsingular at 0.  

\noindent b) For zero and positive energy, the upside-down HO model corresponds to the 
($k = 0$, ordinary) de Sitter and $k = -1$ de Sitter sinh cosmologies,  
with instability causing separation to keep on increasing 
and faster, which motion corresponds to a significant expansion phase in Cosmology. 
For negative energy, it can be taken to bounce ($O_3$) at a finite $\rho_{\tm\ti\tn}$.  
This corresponds to the $k = -1$ de Sitter cosh solution.

\noindent c) This behavior covers pure dust and corresponds to Newtonian gravity mechanics models and 
pure radiation and corresponding to conformal potential mechanics models.
The flat and negatively curved (open in cosmology) models here expand forever ($M_2$) 
while the positively curved  (closed in cosmology) models recollapse ($O_3$) from and to a point 
at which the potential is singular.  

\noindent d) This behaviour covers pure wrong-sign radiation, corresponding to mechanics with 
centrifugal term and/or positive conformal potential.   

\noindent e) This constant-potential case involves expansion forever ($M_2$) 
or everywhere static stable behaviour.

\noindent f) This is wrong-sign radiation (centrifugal term and/or positive conformal potential) 
alongside negative cosmological constant  (alias HO).    

\noindent g) This is wrong-sign radiation corresponding to conformal potential alongside positive 
cosmological constant/HO.    

\noindent h) This is the $\ml \neq$ 0 Hydrogen analogue `Newton-conformal potential 
problem' with oscillatory (`bound') and monotonic (`ionized') regions.  

\noindent i) This is radiation (sufficiently negative conformal potential) alongside negative 
cosmological constant/HO and involves recollapse from and to a point at which the potential is 
singular. 

\noindent j) This is radiation (sufficiently negative conformal potential) alongside positive 
cosmological constant/upside-down HO, covering many different possible behaviours.   

\noindent [Models with dust, radiation and cosmological constant all present \cite{Harrison, 
VajkCG82DS86} exhibit behaviours f), g), i) and j), with scope for turning points in the intermediate 
region.] }        } \label{Fig4}
\end{figure}  }

Some solutions corresponding to QM models studied in this paper are as follows.  
$\rho =  \mbox{sin}(\sqrt{2A}t^{\se\sm})/\sqrt{2A}$ is analogous to the Milne in anti de Sitter 
solution.
$\rho =  \mbox{cosh}(\sqrt{-2A}t^{\se\sm})/\sqrt{-2A}$ is analogous to the positively-curved 
de Sitter model. 
[Both are models with just analogues of $k$, $\Lambda$ of various signs.]
$\rho = (-9K/2)^{1/3}t^{\se\sm\,2/3}$ is analogous to the flat dust model, with well known 
cycloid and hyperbolic counterpart solutions in the positively and negatively-curved cases.  
$\rho = (4(2R - {\cal D}_{\sT\so\st}))^{1/4}t^{\se\sm \, 1/2}$ is analogous to 
the flat radiation model, with well-known curved counterparts (one analogous to the Tolman model). 
This paper's study also requires what is less familiar wrong-sign radiation in cosmology (well-known 
however as ordinary mechanics' centrifugal barrier).  
An example of a solution possessing this is $\rho = \sqrt{t^{\se\sm \, 2} + {\cal D}_{\sT\so\st}}$.

\subsection{Stability of the approximation and sketches of unapproximated potentials}

Throughout, these analogies are subject to any shape factors present being slowly-varying so that one 
can carry out the approximation (\ref{SSA1}--\ref{SSA3}), at least in some region of interest. 
Because of this, one has an analogy between isotropic cosmology solutions and approximate solutions for 
the mechanics of scale. 
E.g. for 3-stop metroland,  
\beq
V = \bar{K}_1|q_2 - q_3|^{\zeta} + \bar{K}_2|q_3 - q_1|^{\zeta} + \bar{K}_3|q_1 - q_2|^{\zeta}
\eeq
(or the sum of various such for different powers $\zeta$).
Then, passing $\rho$, $\varphi$ coordinates 
\beq
V = \bar{K}_1\rho^{\zeta}|\mbox{cos}\,\varphi|^{\zeta} + 
    \bar{K}_2\rho^{\zeta}|\mbox{cos}\,\varphi + \sqrt{3}\mbox{sin}\,\varphi|^{\zeta} + 
    \bar{K}_3\rho^{\zeta}|\mbox{cos}\,\varphi - \sqrt{3}\mbox{sin}\,\varphi|^{\zeta}  \mbox{ } .  
\eeq
This is stable to small angular disturbances about $\varphi$ for some cases of HO/cosmological constant, 
but it is unstable to small angular disturbances for the gravity/dust sign of inverse-power potential.

HO/$\Lambda < 0$ problems include as a subcase finite-minimum wells about the poles of the 4-stop 
metroland configuration space sphere ($C$ = 0 case) \cite{08I, AF, +tri}. 
Adding a ${\cal D}_{\sT\so\st}$ effective term adds spokes to the wells. 
These models are exactly soluble.
Positive power potentials are still finite-minimum wells, but cease to be exactly soluble.    
For Newtonian gravity/dust models (or negative power potentials more generally), near the corresponding 
lines of double collision, the potential has abysses or infinite peaks. 
Thus the scale-dominates-shape approximation is definitely not valid there. 
Thus some assumptions behind the semiclassical approach fail in the region around these lines.  
The upshot is that for negative powers of relative separations the heavy approximation only makes sense in certain 
wedges of angle. 
There is then the possibility that dynamics set up to originally run in such regions leaves them.  
A more detailed stability analysis is required to ascertain whether semiclassicality is stable. 
Fig \ref{Fig5} illustrates this for single and triple potential terms.  
This can be interpreted as a conflict between the procedure used in the semiclassical approach 
and the example of trying to approximate a 3-body problem by a 2-body one.  
See the Conclusion and the Appendix for further comments.

\noindent 

{            \begin{figure}[ht]
\centering
\includegraphics[width=0.5\textwidth]{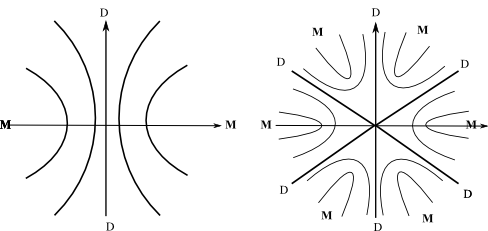}
\caption[Text der im Bilderverzeichnis auftaucht]{        \footnotesize{Contours for single and triple 
negative power potentials. 
These have abysses along the corresponding double collision lines and high ground in between them (for 
negative-power coefficients such as for the attractive Newtonian gravity potential).  
}        }
\label{Fig5}
\end{figure}  }

\section{Setting up QM for scaled metroland models}

\subsection{Kinematical quantization}

Kinematical quantization is an important preliminary step in which quantites appropriate to use at the 
quantum level are selected -- a minimal-sized vector space $V$ and the canonical group of the system, 
${\cal G}$.
(For a more general treatment than this paper's example below, and a number 
of associated subtleties, see \cite{I84}).  
For the kinematical quantization of scaled $N$-stop metroland, select the following quantities. 

\noindent 
1) n suitable periodic configuration variables $n^i$ that lie on the hypersphere form $V$

\noindent
2) n ordinary momenta (corresponding to the translations). 

\noindent
3)(N -- 1)(N -- 2)/2 $SO$(N -- 1) objects ${\cal D}_{\Delta}$ (corresponding to the rotations).  
The canonical group ${\cal G}$ here is just  the $n$-dimensional Euclidean group, Eucl($n$), and the 
final relevant structure is $V \mbox{\textcircled{S}} {\cal G} = \mathbb{R}^{n} \mbox{\textcircled{S}} 
\mbox{Eucl}(n)$, where $\mbox{\textcircled{S}}$ denotes semi-direct product.

\subsection{Time-independent Schr\"{o}dinger equations}

Use the cone structure of the shape-scale split form of the metric \cite{Cones}, the Laplacian on 
relational space is $D^2\Psi = 
\frac{1}{\sqrt{\cal M}}
\frac{\pa}{\pa {\cal R}^{p}}\left(\sqrt{{\cal M}}{\cal N}^{pq}\frac{\pa}{\pa {\cal R}^{q}} 
\right) = 
\rho^{1 - nd}\pa_{\rho}(\rho^{nd - 1}\pa_{\rho}\Psi) + \rho^{-2}D_{S}^2\Psi$ 
(for the Laplacian on shape space ${D}_{S}^2 =$  

\noindent$\frac{1}{\sqrt{M}}\frac{\pa}{\pa S^{u}}\left(\sqrt{M}N^{uv}
\frac{\pa}{\pa S^{v}}\right)$) and general $\xi$-operator-ordering 
\cite{Ordering, HP86, Wiltshire}.  
Thus ${\cal N}^{pq}{\cal P}_{p}{\cal P}_{q}\longrightarrow D^2 - \xi\,\mbox{Ric}(M)$ a 
general time-dependent Schr\"{o}dinger equation for scaled $N$-stop metroland or $N$-a-gonland is  
\beq
-\hbar^2(\rho^{1 - nd}\pa_{\rho}(\rho^{nd - 1}\pa_{\rho}\Psi) + \rho^{-2}
(D_{S}^2  \Psi - \kappa(\xi)\Psi)) + 2V(\rho, S^{u})\Psi = 2E\Psi 
\mbox{ } .  
\eeq
Here, $\kappa(\xi)$ is a constant equal to $\xi\,\rho^2\,\mbox{Ric}(M)$. 
This is 0 for the cone C($\mathbb{S}^{n - 1}$) = $\mathbb{R}^{n}$ with flat metric.  
Thus there is no Laplace/conformal/$\xi$ operator ordering distinction for scaled $N$-stop metroland. 
On the other hand, it is $6\,n\,\xi$ for the $N$-a-gonland relational spaces which are the cones 
C($\mathbb{CP}^{n - 1}$), which, in the conformal-ordered case, gives $3n(2n - 3)/4(n - 1)$.  
(Conformal ordering is motivated by being the member of the family of orderings invariant under 
changes of coordinatization on the configuration space that furthermore preserves a conformal 
invariance that can be traced back to the form of the relational action \cite{Banal}.)
The absence of this ambiguity in the present paper's scaled $N$-stop metroland paper saves us quite a 
lot of work. 
However, this absence is also a limitation since this ambiguity is indeed of interest as regards 1) the 
operator ordering problem in Quantum Cosmology and Quantum Gravity. 
2) The absolute versus relative motion debate \cite{Cones}.   
The absence of this complication is one good reason to study $N$-stop metroland a a precursor to 
studying triangleland \cite{08III, FileR} and $N$-a-gonland.  
Another good reason for studying this first is that it does not distinguish between reduced and Dirac 
quantization schemes, so that a second source of ambiguities is also absent.

\noindent For the particular cases of this paper, $d$ = 1 has $D_{S}^2 = 
D_{\mathbb{S}^2}^2$, and then, more specifically for $N$ = 3 and 4, 

\noindent 1) for 3-stop metroland, using the conformal (and, as flat, equivalently the elementarily 
familiar Laplacian) ordering, 
\beq
-\frac{\hbar^2}{2}
\left(
\frac{1}{\rho}\frac{\pa}{\pa\rho}\left(\rho\frac{\pa}{\pa\rho}\right) + 
\frac{1}{\rho^2}\frac{\pa^2}{\pa\varphi^2}
\right)\Psi 
+ V(\rho, \varphi)\Psi = E\Psi 
\label{TISE3}
\eeq

\noindent
2) Likewise, for 4-stop metroland, using the conformal (and, as flat, equivalently the elementarily 
familiar Laplacian) ordering,
\beq
-\frac{\hbar^2}{2}
\left(
\frac{1}{\rho^2} \frac{\pa}{\pa\rho} 
\left(
\rho^2\frac{\pa}{\pa\rho}
\right) 
+ \frac{1}{\rho^2\mbox{sin}\,\theta}\frac{\pa}{\pa\theta}
\left(
\mbox{sin}\,\theta\frac{\pa}{\pa\theta}
\right) 
+ \frac{1}{\rho^2\mbox{sin}^2\theta} \frac{\pa^2}{\pa\phi^2}
\right)\Psi + V(\rho, \theta, \phi) \Psi = E\Psi
\label{TISE4}
\eeq


\noindent
The following separation ansatz is useful below: 
\beq
\Psi = F(\rho)Y(\theta_{r}) \mbox{ } .
\label{sssplit}
\eeq

\section{Simple quantum solutions in the scale-dominates-shape approximation}

\subsection{The common separated-out shape part}

This gives in general the hyperspherical harmonics equation \cite{v2}, the 2- and 3-d cases 
corresponding to 3-stop and 4-stop metroland are then of very well known mathematical forms. 
For 3-stop metroland, we use a `D-M'choice of basis with the pair of D's corresponding to the (1) 
cluster providing the principal axis and the pair of M's perpendicular to it providing the second axis.  
The solutions are 
\beq
Y_{\sd}(\varphi) \propto \mbox{exp}(i \d \varphi) \mbox{ } ;
\label{16}
\eeq
one would often then take sine and cosine combinations of these.  
Then in terms of shape quantities, 
\beq
Y_{\sd} \propto{\cal T}_{\sd}(n_1) \mbox{ } ,  
\eeq
i.e. a function of $n_1$ alone, which, we remind the reader, is a measure of how large cluster 1b is 
relative to the size of the model universe.   
Here, ${\cal T}_{\sd}(X)$ is defined to be $T_{\sd}(X)$ for cosine solutions and $\sqrt{1 - T_{\sd}(X)^2}$ 
for sine solutions, where $T_{\sd}(X)$ is the Tchebychev polynomial of degree d.
Also, d $\in \mathbb{Z}$ is a total relative dilational quantum number.

Interpretation of the first few solutions is as follows.  
For d = 0, one has the same probability for all ratios of $\rho_1$ and $\rho_2$ 
(for any clustering). 
For the d = 1 cosine solution, cluster 23 being small compared to its separation from particle 1 is 
favoured (i.e. a well-separated model universe). 
On the other hand, for the d = 1 sine solution, cluster 23 being large compared to its separation from 
particle 1 is favoured (i.e. a well-merged model universe).
For the d = 2 cosine solution, both of the preceding are equally favoured with gaps in between, 
while for the d = 2 sine solution both of the preceding are equally disfavoured with peaks in between.  
For the d = 3 cosine solution, near-D configurations for all six possible D's are favoured with near-M 
configurations for all 6 possible M's being disfavoured, and vice versa for the d = 3 sine solution.

{            \begin{figure}[ht]
\centering
\includegraphics[width=0.6\textwidth]{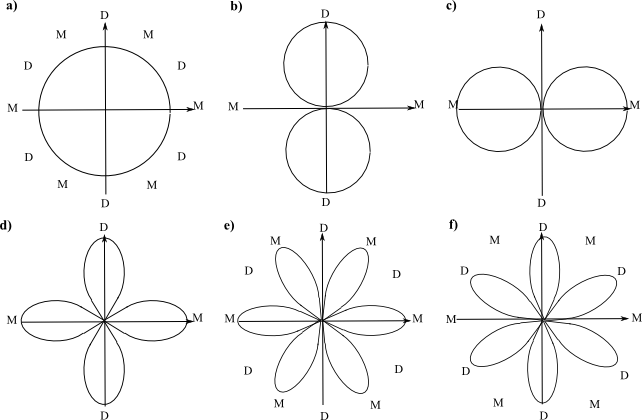}
\caption[Text der im Bilderverzeichnis auftaucht]{        \footnotesize{Separated-out shape part for 
3-stop metroland.}   }
\label{Fig6} 
\end{figure}  }

For 4-stop metroland, we have  
\beq
Y_{\sD\sd}(\theta, \phi) \propto \mbox{P}_{\sD}^{\sd}(\mbox{cos}\,\theta)\mbox{exp}(i\d \phi) 
\label{Nye} \mbox{ } .  
\eeq
Here, $\mP_{\sD}^{\sd}(\xi)$ are associated Legendre functions.
Again, one would often then take sine and cosine combinations of these.  
Then in terms of shape quantities \cite{AF}, 
\beq
Y_{\sD\sd} \propto \mbox{P}_{\sD}^{\sd}(n_3)
{\cal T}_{\sd}(n_1/\sqrt{1 - n_3\mbox{}^2}) \mbox{ } .  
\eeq
I.e., a function of $n_3$: the separation of cluster 12 from cluster 23 relative to the size of the 
model universe, multiplied by a function of $n_3$ and $n_1$: the size of cluster 12 relative to the 
size of the model universe.
Here, D $\in \mathbb{N}_0$ and $d \in \mathbb{Z}$ subject to $|\md| \leq \mD$ are `total' and 
`projected' relative dilation quantum numbers.
The corresponding analysis for the 4-stop case that applies throughout this Sec coincides with that 
already presented in \cite{AF} by the shape--scale split.

\subsection{Analogues of $k < 0$ vacuum or wrong-sign radiation}

These are free models with zero and nonzero total dilational quantum number respectively.  
They require $E > 0$ (i.e. the analogue of negative curvature $k < 0$) in order to be physically 
realized (elsewise $E$ lies on or below the potential everywhere, so there is no solution).  
Then one has solutions in terms of Bessel functions, 
\beq
F_{\sd}(\rho) \propto \mJ_{\sd}(\sqrt{2E}\rho/\hbar)   
\eeq
for 3-stop metroland, and, for 4-stop metroland, 
\beq
F_{\sD}(\rho) \propto \rho^{-1/2}\mJ_{\sD + 1/2}(\sqrt{2E}\rho/\hbar)
\eeq
(i.e. spherical Bessel functions).  
Each case's probability density function exhibits an infinity of oscillations in the scale direction.  
However, free models have limitations through unboundedness/non-normalizability, and, also, as regards 
setting up the semiclassical approach, through being exactly separable into shape and scale parts. 
Thus we also need to consider cases with nontrivial potentials.

\subsection{Analogues of $\Lambda < 0$, $k < 0$ vacuum or wrong-sign radiation}

For 3-stop metroland with $V = A\rho^2$, $A > 0$, (\ref{TISE3}) gives the 2-$d$ isotropic quantum HO 
problem \cite{Robinett, Schwinger} (but now in {\sl configuration space}).   
In shape-scale variables, the scale part is {\sl exactly} solved by \cite{MGM} (\ref{sssplit}, \ref{16}) and 
\beq
F_{\sN\,\d}(\rho) \propto \rho^{|\d|}\mL_{\sN}^{\sd}(\omega\rho^2/\hbar)\mbox{exp}(-\omega\rho^2/2\hbar)
\eeq
for $\mL^{\sd}_{\sN}(\xi)$ the generalized Laguerre polynomials and $\omega = \sqrt{2A}$.  
These solutions correspond to energies 

\noindent $E = \hbar\omega(2\mN + |\md| + 1) > 0$ for 
$\mN \mbox{ } \in \mbox{ } \mathbb{N}_0$ and $\d \mbox{ } \in \mbox{ } \mathbb{Z}$.  
These wavefunctions are finite at 0 and $\infty$ in scale, with N nodes in between.

For 4-stop metroland with $V = A\rho^2$, (\ref{TISE4}) gives the 3-$d$ isotropic quantum HO problem 
(in configuration space). 
In shape-scale variables, the scale part of this is solved by (\ref{sssplit}, \ref{Nye})  
\beq
F_{\sN\,\sD}(\rho) \propto \rho^{\sD}\mL_{\sN}^{\sD + 1/2}(\omega\rho^2/\hbar)
\mbox{exp}(-\omega\rho^2/2\hbar)
\mbox{ } ,
\eeq
corresponding to energies $E = \hbar\omega(2\mN + \mD + 3/2) > 0$ for $\mN \in \mbox{ } \mathbb{N}_0$. 
Again, these wavefunctions are finite at 0 and $\infty$ in scale, with $N$ nodes in between.

\subsection{Analogues of $\Lambda > 0$, $k < 0$ vacuum models} 

For $V = A\rho^2$, $A < 0$, (\ref{TISE3}) and (\ref{TISE4}) give instead isotropic upside-down HO 
mathematics (see e.g. \cite{Robinett}).    
The solutions of these are oscillatory rather than tightly localized.

\subsection{Analogues of $k > 0$ dust model}

The 4-stop and thus $\mathbb{R}^3$ configuration space case of this is mathematically well known 
(e.g. \cite{LLQM}): it is the analogue of the l = 0 hydrogen model, 
\beq
F_{\sN}(\rho) \propto \mL_{\sN - 1}^{1}(2\sqrt{-2E}\rho/\hbar)\mbox{exp}(\sqrt{-2E}\rho/\hbar)
\mbox{ } .  
\eeq
The working involved also fairly straightforwardly generalizes to the 3- and $N$-stop 
counterparts.  
For the 3-stop case, 
\beq
F_{\sN}(\rho) \propto \mL_{\sN - 1 }(2\sqrt{-2E}\rho/\hbar)\mbox{exp}(\sqrt{-2E}\rho/\hbar) \mbox{ } ,
\eeq
corresponding to energies $E = - k^2/2\hbar^2(\mN - 1/2)^2$.
%
%
Nor is it hard to extend the above to also include the wrong-sign radiation case ${\cal D}_{\sT\so\st} 
> 0$.

\subsection{Analogues of $k \leq 0$ dust models}

These are `ionized states' or `scattering problems' and correspond to open cosmologies 
The mathematics of the ionized atom (continuous spectrum part) can be found in e.g. \cite{LLQM, CH}.

\subsection{Analogues of right and wrong sign radiation models}

Again, for these potentials, 4-stop metroland's $\mathbb{R}^3$ configuration space mathematics is 
well-known \cite{LLQM}.  
It has three cases, 
i) $\mD + 1/2 < \sqrt{2R}/\hbar$ for which there is collapse to the maximal collision (ground 
state with $E = - \infty$). 
This is relevant since the approximate 
problem's scale part can exchange  energy with the shape problem 
and thus use this energy exchange to run down to the maximal collision. 
ii) $\mD + 1/2 > \sqrt{2R}/\hbar$ (for which it is not clear that the cutoff in \cite{LLQM}
is meaningful in the case of RPM's). 
iii) The critical case $\mD + 1/2 = \sqrt{2R}/\hbar$ has $F$ diverge no worse than 
$1/\sqrt{\rho}$ as $\rho \longrightarrow 0$.

The 2-$d$ configuration space case of 3-stop metroland is somewhat different: 
i) $\md < \sqrt{2R}/\hbar$ is still a collapse to the maximal collision. However, 
ii) $\md  > \sqrt{2R}/\hbar$ now has a positive root choice, which is finite as 
$\rho \longrightarrow 0$.
iii) The critical case $\md = \sqrt{2R}/\hbar$ is now also finite as $\rho \longrightarrow 0$.

\subsection{Approximate $\rho$ behaviours at the quantum level}

In summary, each $O_1$ is bound in an interval including 0 and each $O_2$ is bound in an interval 
excluding (shielded from) 0.  
$O_3$, $M_1$, $M_2$ are all `tending to free behaviour at large $\rho$. 
Forbidden regions exhibit exponential decay. 
Finally, the $O_2$--forbidden-region--$O_3$ and $A_1$--$S_1$--$A_2$ slices become tunnelling scenarios.

\section{The full quantum problem}

\subsection{The general quadratic potential in 3-stop metroland}

In $\rho_1, \rho_2$ variables, this is also solved standardly.  
It is extendible to obtain the {\sl exact} solution of the $B \neq 0$ case (i.e. non-equal 
mass-weighted Hooke's coefficients, $K_1 \neq K_2$) -- for energies $E = \hbar\sqrt{K_1}(\mn_1 + 1/2) + 
\hbar\sqrt{K_2}(\mn_2 + 1/2)$.  
Here, $\mn_1, \mn_2 \mbox{ } \in \mbox{ } \mathbb{N}_0$.  
Next, passing to shape-scale variables, the wavefunctions are  
\beq
\Psi_{\sn_1\sn_2}(\rho, \varphi) \propto 
\mH_{\sn_1}\big(\sqrt{{K_1}/{\hbar}}\rho\,\mbox{cos}\,\varphi\big) 
\mH_{\sn_2}\big(\sqrt{{K_1}/{\hbar}}\rho\,\mbox{sin}\,\varphi\big)
\mbox{exp}\big({-\rho^2(A + B\,\mbox{cos}\,2\varphi)}/{\hbar}\big) \mbox{ } ,  
\eeq
where the $\mH$'s denote Hermite polynomials.  
One can think of these wavefunctions as rectangular arrays of peaks and troughs (note that peak 
strength is not constant).
The ground state favours all D's and M's equally.
Then in the D--M basis, one of first excited states favours the pair of D's and disfavours the pair of 
M's and vice versa for the other excited state; the maximal collision O is also disfavoured in each case.
Second excited states with two 0 quantum numbers and a 2 are likewise but with three lobes, so that O is 
also favoured while two nonzero distances have disfavoured nodes.  
On the other hand, the second excited state with two 1 quantum numbers disfavours all four of these D's 
and M's. 
The $B \neq 0$ case (for $|B| < A$) is then a distortion of this (stretched in one direction and 
squeezed in the perpendicular direction).

Also, the potential $U(1 - \mbox{cos}\,2\theta)$ for $U$ constant occurs in modelling the 
rotation of a linear molecule in a crystal \cite{P,WSPW}.  
In the 3-stop metroland case, this gives 
\cite{P} the fairly standard mathematical physics of Mathieu's equation \cite{AS}.

\subsection{The special quadratic potential in 4-stop metroland}

In $\rho_1, \rho_2, \rho_3$ variables, it is solved in terms of Hermite polynomials and Gaussians, 
and this extends to $B \neq 0$ case as well, which is, likewise, a distortion.  
For energies $E = \sum_{e = 1}^{3} \hbar\sqrt{K_e}(\mn_e + 1/2)$ ($k_e = \omega_e/\hbar$ and 
$\mn_e \mbox{ } \in \mbox{ } \mathbb{N}_0$, this is solved standardly in the associated Cartesian 
coordinates $\rho_e$ in Hermite polynomials $H_{n_e}$. 
I then express the arguments of these in terms of the scale variable $\rho$ and the shape variable 
$\theta$, $\phi$ to obtain 
\beq
\Psi_{\sn_1\sn_2}(\rho, \theta, \pi) \propto 
\mH_{\sn_1}\big(\sqrt{\mbox{$\frac{K_1}{\hbar}$}}\,\rho\,\mbox{sin}\,\theta\,\mbox{cos}\,\phi\big) 
\mH_{\sn_2}\big(\sqrt{\mbox{$\frac{K_1}{\hbar}$}}\,\rho\,\mbox{sin}\,\theta\,\mbox{sin}\,\phi\big)
\mH_{\sn_3}\big(\sqrt{\mbox{$\frac{K_1}{\hbar}$}}\,\rho\,\mbox{cos}\,\theta\big)
\mbox{exp}\big(\mbox{$\frac{-\rho^2(A + B\,\mbox{cos}\,2\theta + C\,\mbox{sin}^2\theta \,\mbox{cos}\,2\phi)}{\hbar}$}\big)
\eeq
\mbox{ } \mbox{ } One can likewise think of these as box-shaped arrays of peaks and troughs.  
The ground state favours all DD's  and T's equally.
Then in the `DD' basis whose principal axis is centred about the \{12, 34\} DD collision, one first 
excited state favours this DD pair while also disfavouring O.  
The other two excited states favour each of the other two DD pairs.  
The second excited states with two quantum numbers zero and the remaining one a 2 are likewise 
but consisting of 3 lobes, so that O is also favoured while two nonzero distances have disfavoured nodes.  
Those with two quantum numbers 1 and the remaining one zero disfavour all DD's while somewhat 
favouring the T's.  
The third excited state includes the state with all quantum numbers taking the value 1.  
This picks out all 8 T's while all the DD's are nodal.  
The $B \neq 0$ case (for $|B| < A$) is again a distortion of this (stretched in one direction and 
squeezed in a perpendicular direction).

\subsection{Note about other cases}

An approximate solution for e.g. the ground state of the isotropic upside-down HO in 3-stop metroland is 
(using separation in Cartesian coordinates \cite{Robinett} and then changing to shape--scale variables)
\beq
\Psi \propto 
\mbox{sin}({\sqrt{-K_1/\hbar}\rho^2\mbox{cos}^2\varphi}/{2})
\mbox{sin}({\sqrt{-K_2/\hbar}\rho^2\mbox{sin}^2\varphi}/{2})
/\rho\sqrt{\mbox{sin}\,2\varphi} \mbox{ } .  
\eeq
The probability density function for this is a rectangular grid of humps that decrease in height as one 
moves out radially.
Getting angular dependence via $B \neq 0$ continues to be straightforward in this case, readily 
allowing study of models for which the premises of the semiclassical approach apply.  
This is also the case for what is mechanically a mixture of HO's and upside-down HO's.
%
%
For other power-law potentials, further approximate or numerical work is required.

\section{Interpretation of the solutions.}

\subsection{Characteristic scales}

RPM models with $K/\rho$ potentials have a `Bohr moment of inertia' (square of `Bohr hyperradius') for 
the model universe analogue to (atomic Bohr radius)$^2$, $I_0 = \rho_0^2$. 
In the gravitational case, this goes as $\hbar^4/G^2m^5$.  
Then $E = - \hbar^2/2I_0(\mN - 1/2)^2$ for $N = 3$ and $E = - \hbar^2/2I_0\mN^2$ for $N = 4$.
HO RPM models have characteristic $I_{\sH\sO} = \hbar/\omega$.  
Note that the `Bohr' case is limited by the breakdown of the approximations used in its derivation,  
while the HO case has no such problems.

\subsection{Expectations and spreads in analogy with molecular physics}

Expectations and spreads of powers of the radial variable $r$ are used in the study of atoms. 
(See e.g. \cite{Messiah} for elementary use in the study of hydrogen, or \cite{FF} for use in 
approximate studies of larger atoms).   
These provide distinct information about the probability distribution function from that obtained from 
the `modal' quantities (peaks and valleys) that are read off from plots or obtained using calculus.   
E.g. for hydrogen, one obtains (e.g. in the $\ml = 0$ case)
\beq
\mbox{expectation} = \langle\mn\,0\,0\,|\,r\,|\,\mn\,0\,0 \rangle = 3\mn^2a_0/2 
\mbox{ } \mbox{ and } \mbox{ } 
\mbox{spread} = \Delta_{\sn\,0\,0}r = \sqrt{\mn^2(\mn^2 + 2)}a_0/2
\eeq
where $a_0$ is the Bohr radius of the atom.
%
%
One can then infer from this that a minimal typical size is $3a_0/2$ and that the radius and its spread 
both become large for large quantum numbers.  
Contrast with how the modal estimate of minimal typical size is $a_0$ itself.  
The slight disagreement between these serves as some indication of the limited accuracy to which either 
estimate should be trusted.    
Also, the above can be identified as expectations of scale operators, and thus one can next ask 
whether they have pure shape counterparts in the standard atomic context.    

\mbox{ }

\noindent 
{\bf Example 1)} the most direct counterpart of this atomic analogy is the D = 0 approximate Newtonian 
gravity or attractive Coulomb problem, which is the $a_0 \longrightarrow \rho_0$ of the preceding.  

\mbox{ } 

\noindent 
{\bf Example 2)}  As regards the expectation of the size operator $\widehat{\rho}$ for 3-stop metroland 
isotropic case in shape-scale coordinates, $\langle\,0\,\d\,|\widehat{\rho}|\,0\,\d\,\rangle = 
\mbox{\Large(}
\stackrel{2|\d| + 1}{\mbox{\scriptsize{$|\d|$}}}
\mbox{\Large)}
\frac{\sqrt{\pi}(|\d| + 1)}{2^{2|\td| + 1}}\rho_{\sH\sO}$ \mbox{ } . 
%
%
Thus e.g. $\langle\,0\,0\,|\widehat{\rho}|\,0\,0\,\rangle = \frac{\sqrt{\pi}}{2}\rho_{\sH\sO}$ 
and the large-$|\d|$ limit for N = 0 is $\sqrt{{\hbar|\d|}/{\omega}}$.  
The latter slowly rises to become arbitrarily large for configurations possessing more and more relative 
dilational momentum.   
Also, $\langle\, 1\, 0\,| \widehat{\rho} | \,1\,0\, \rangle = \frac{7\sqrt{\pi}}{8}\rho_{\sH\sO}$. 
In comparison, the modal value for the ground state is $\rho_{\sH\sO}/\sqrt{2}$.

The spreads in size are an integral made easy by a recurrence relation on the generalized Laguerre 
polynomials, $\langle\, \mN\,\d\,| \widehat{\rho}\mbox{}^2 | \,\mN\,\d\, \rangle$ 
$= (2\mN + |\d| + 1)\rho_{\sH\sO}\mbox{}^2$, minus the square of the expectation.  
Thus $\Delta_{0\,0}(\widehat{\rho}) = \frac{4 - \pi}{4}\rho_{\sH\sO}\mbox{}^2 \approx 0.215 
\rho_{\sH\sO}\mbox{}^2$, while the large-$|\d|$ limit with N = 0 gives 
$\Delta_{0\,|\sd|}(\widehat{\rho}) \longrightarrow \rho_{\sH\sO}\mbox{}^{2}$, and 
$\Delta_{1\,0}(\widehat{\rho}) = \frac{192 - 49\pi}{64}\rho_{\sH\sO}\mbox{}^2 \approx 0.595 
\rho_{\sH\sO}\mbox{}^2$.

The expectations of the 3-stop metroland shape operators are all zero by two-angle formulae and the 
orthogonality of Fourier modes. 
Likewise, the spreads of the shape operators are 1/2 in all states.  
All of the above calculations benefit from factorization into shape and scale parts, with insertion of a  
pure-shape operator rendering the scale factor trivial and vice versa.

\mbox{ } 

\noindent{\bf Example 3)} as regards the expectation of the size operator for 4-stop metroland, 
isotropic case in shape-scale coordinates, 

\noindent
%
$\langle \, 0\,\mD\,\d\,| \widehat{\rho} | \,0\,\mD\,\d\, \rangle = 
\frac{2^{2\tD + 2}}{\sqrt{\pi}}
\mbox{\Large(}
\stackrel{2\sD + 2}{\mbox{\scriptsize $\sD + 1$}}
\mbox{\Large)}^{-1} \rho_{\sH\sO}$. 
Thus e.g. 
$\langle\,0\,0\,0\,|\widehat{\rho}|\,0\,0\,0\,\rangle = \frac{1}{2\sqrt{\pi}}\rho_{\sH\sO}$.  
Also, the large D limit (for $\mN =  0$) is 
$\langle\,0\,\mD\,\d\,|\widehat{\rho}|\,0\,\mD\,\d\,\rangle$ $\longrightarrow \sqrt{\mD + 1}
\rho_{\sH\sO}$. 
This again gradually grows to infinity along a sequence of configurations with ever-increasing relative 
dilational momentum.  
Also, $\langle\,1\,0\,0\,|\widehat{\rho}|\,1\,0\,0\,\rangle = \frac{3}{\sqrt{\pi}}\rho_{\sH\sO} 
\approx 1.69\rho_{\sH\sO}$.   
In comparison, the modal value for the ground state is $\rho_{\sH\sO}$.

The spreads are again an integral made easy by a recurrence relation on the associated Laguerre 
polynomials, $\langle\, \mN\,\mD\,\d| \widehat{\rho}^2 | \,\mN\,\mD\,\d\, \rangle = (2\mN + \mD 
+ 3/2)\rho_{\sH\sO}\mbox{}^2$, minus the square of the expectation.  
Thus $\Delta_{0\,0\,0}(\widehat{\rho}) =  \frac{6\pi - 1}{4\pi}\rho_{\sH\sO}\mbox{}^2 \approx 
1.42\rho_{\sH\sO}\mbox{}^2$, while the large-$\mD$ limit with N = 0 gives 
$\Delta_{0\,\sD\,\sd}(\widehat{\rho}) \longrightarrow \rho_{\sH\sO}\mbox{}^2/{2}$, and 
$\Delta_{1\,0\,0}(\widehat{\rho}) = \frac{7\pi - 18}{2\pi}\rho_{\sH\sO} \approx 0.635 \rho_{\sH\sO}$.

For expectations and spreads of the 4-stop metroland shape operators, \cite{AF} gives these immediately 
by the shape--scale split.  
Furthermore these overlap integrals are of 3-$Y$ integral form.  
For these, the most general case is known explicitly \cite{LLQM} in terms of what are, mathematically, 
Wigner 3j symbols' (though for us, physically, they are `3d symbols').  
The specific cases relevant to 4-stop metroland with harmonic oscillator-like potentials (where the 
sandwiched $Y$ is of degree 2) are explicitly written out in e.g. \cite{Mizu}.  
See also SSecs 8.2, 8.4 and 8.5 for further interpretative points.

\subsection{Further perturbative treatment} 

The cosmology--RPM analogy in this paper gives multiple power-law potential terms; furthermore some 
of them admit sensible expansions following from the scale-dominates-shape approximation.
In a number of cases some such terms can be regarded as small.  
The cases in which the scale-dominates-shape approximation holds are of interest partly through being  
one of the conditions behind arriving at a long-term-stable semiclassical regime.
The negative power cases are not expected to work well already from the classical analysis.  
This is because, as further discussed in the Conclusion, shape is simply not secondary in some parts of configuration 
space in such cases. 
[Also, there are extra layers of control for HO's: one can set these up so that semiclassicality applies  
everywhere. 
One can then solve exactly even in cases for which there is not separability in the $h$--$l$ variables.]

Next, I note that the above exact solution work can be perturbed about, with a number of cases of this 
producing standard mathematics as follows.   
The perturbation integrals split into scale parts and shape parts.  
The scale integrals give $\langle\rho^{\alpha}\rangle$ for $\alpha$ an integer (taken to be positive if one is to 
avoid classical instability).  
Such integrals are of the same type as those used in the evaluation of expectations and spreads in Sec 
7.2 (and so are adaptable from results in e.g. \cite{Messiah}). 
The shape integrals for 4-stop metroland involve $\phi Y_{\sD\sd}(\theta,\phi)Y_{\sD\sd}(\theta,\phi)$ 
and $\theta Y_{\sD\sd}(\theta,\phi)Y_{\sD\sd}(\theta,\phi)$. 
The former is trivial and the latter is straightforward, at least state-by-state.  
For 3-stop metroland, these are of the form $\varphi\, \mbox{exp}(i \d \varphi)$, and thus trivial again.  
In \cite{AF} instead we treated the further HO terms not by scale-dominates-shape approximation expansion but 
as an exact perturbation.

\section{Conclusion}

Relational particle models (RPM's) are useful toy models for investigating the Problem of Time and 
other foundational aspects of Quantum Cosmology.  
This is due to their resemblance to the geometrodynamical formulation of GR and in ways distinct to 
minisuperspace models (including nontrivial notions of inhomogeneity/structure).  
The present paper extends \cite{AF}'s spatially 1-$d$ models to include scale,  
which is essential for a closer analogy with Cosmology.
It also extends \cite{MGM} from 3 to 4 and $N$ \cite{v2} particles. 
It is furthermore a useful precursor for 1) the more complicated case of the scaled triangle formed by 3 
particles in 2-$d$ which is work in progress \cite{08III, FileR}. 
2) For subsequent work on the even more interesting (while still just about modellable) case of the 
quadrilateral formed by 4 particles in 2-$d$ \cite{QShape}.  
In the present paper, scaled RPM's freedom in the choice of potential is partly used up by the 
Cosmology--Mechanics analogy (see also \cite{Cones}). 
Here, the hyperradius $\rho$ (square root of the moment of inertia) plays an analogous role to the 
cosmological scalefactor $a$. 
Also, --2 times the energy plays an analogous role to the spatial curvature.

Particular features of models following from this that are studied in the present paper are 
i) multi (upside-down) HO's without or with total dilational momentum ${\cal D}_{\sT\so\st}$. 
This corresponds to models with cosmological constant and without or with wrong-sign radiation. 
ii) Newtonian gravity potentials (approximated at least in some regions by dust terms).  
iii) conformal potentials (inverse square potentials, approximated at least in some regions by radiation 
terms).

The present paper can be seen as applying further filters to this set of geometrodynamically and 
(quantum) cosmologically inspired toy models. 
A first such filter is the stability of the scale-dominates-shape approximation. 
This is a particular example of some of the approximations underlying the semiclassical approach. 
Further such filters are boundedness and other good behaviour of the wavefunction, 
and mathematical tractability of the QM.

I have provided a number of exact and approximate quantum solutions for the above 
cosmologically-inspired potential choices of RPM models.  
In interpreting these, I have noted that some possess a characteristic scale: a `Bohr moment of inertia  
of the model universe' analogue of the Bohr radius of the atom and an oscillator scale.
I have also considered expectations and spreads for these quantum solutions and the tractability of 
further perturbations about them.

Some Problem of Time applications and further quantum cosmological interpretation of 
the present paper are given in the remaining SSecs of the Conclusion.
All of these applications (some of which are to be presented in more detail elsewhere \cite{FileR, 
SemiclIII, FileR}) remains a useful precursor for further triangleland and quadrilateralland toy model work 
\cite{08III, SemiclIII, FileR}.\footnote{The triangleland model also has spherical and $\mathbb{R}^3$ 
mathematics, albeit the interpretation of this in terms of geometry and physics is rather more 
complicated than in the present paper's 4-stop metroland case.
The quadrilateral model has the mathematics of $\mathbb{CP}^2$ and the cone over that.}  
These models have further useful features as regards analogy with GR Quantum Cosmology beyond 
those analogies in the present paper's simpler models. 
For example, they possess linear constraints in addition to the notions of inhomogeneity (particle 
clumping) that the present paper's $N$-stop metrolands already have.
The quadrilateral case additionally possesses less trivial notions of subsystem, which is relevant 
to the study of the below-mentioned timeless approaches.
As such it is better to mostly wait for these more advanced models as regards most details of 
applications \cite{SemiclIII, FileR}, rather than to press on with these in the present paper's context 
(a few brief points that the present paper's models already illustrate are made below).  
The current paper's ideas -- of using up freedom in the potential for the scale part of the system 
to mimic GR cosmological solutions, thus establishing a GR-RPM analogy but now with a geometrically 
simpler and finite notion of inhomogeneity coupled to this and therefore allowing for much more 
straightforward qualitative investigation of the Halliwell--Hawking semiclassical approach to the 
quantum-cosmological origin of structure formation -- carry over to these more advanced models. 
(In general, the form of the analogy is the same for all models, excepting the spherical realization 
of the triangleland model, which differs in this respect through using $I = \rho^2$ and not $\rho$ 
as its scale variable; this notable difference is presented in \cite{08III}.)

\subsection{Emergent semiclassical time approach for $N$-stop metrolands}

The present paper's work permits generalization of \cite{MGM}'s semiclassical approach for 3-stop 
metroland with HO-like potentials (this is further developed in \cite{SemiclIII}).   
In this approach, the scale is the heavy slow $h$ degree of freedom and the pure shape is made up of 
light, fast $l$ degrees of freedom.
The present paper adds to this by providing an example of how some semiclassical approach approximations 
only hold in some regions.
Namely, that some of these manifest themselves as the scale-dominates-shape approximation (\ref{SSA1}, 
\ref{SSA2}), about which we have established the following.

On the one hand, HO/cosmological constant models are especially well-behaved in this respect. 
Here, the scale-dominates-shape approximation is enforceable by suitable choice of potential 
coefficients $|B/A|$, $|C/A|$ small.  
(This corresponds to a high but not perfect homogeneity of the contents of the universe, corresponding 
to there being similar but not identical Hooke's coefficients for the `springs' between particles.)

\noindent 0)
On the other hand, 3-stop metroland with negative power law potentials has problems with this 
approximation. 
This is because it only holds well in some wedges of configuration space, and 
trajectories/wavepackets are then capable of leaving these regions.
These problems are of a kind that is well-known mathematically, through their being closely related to 
how the 3-body problem is often not well-approximated by the 2-body problem. 
The present paper finds a context in which this mathematical problem applies in the context of a 
semiclassical approach to a toy model universe quantum cosmology.   
Moreover, there is also a further and distinct Cosmology--RPM analogy at the level of the Hamiltonian. 
(This follows through to the fully-quantum counterpart of the Hamiltonian and are presented in the 
Appendix). 
Then, by this distinct analogy, GR quantum cosmologies at least usually have positive power potentials. 
Examples of such positivity are \cite{HH83} (scalar field and cosmological constant terms), 
\cite{Amsterdamski, FangMo} (anisotropies) and in the approximate treatment of small inhomogenities 
\cite{HallHaw}. 
Thus the difficulty encountered in the present paper -- of inverse power potentials making neglecting 
non-size degrees of freedom an unstable approximation -- is at least usually absent from GR Quantum 
Cosmology.  
Thus I view this example as but a mild reason to question some of the semiclassical approach's 
approximations. 
More generally, I note that the present paper's tessellation method facilitates the determination of in 
which regions various semiclassical approximations hold in RPM's.

The approximate emergent time $t^{\se\sm}$ that the $h$-system provides is computed for RPM's in 
\cite{Cones}. 
For the five classical solutions in Sec 3.4, respectively, one gets
$t^{\se\sm} = \mbox{arcsin}(\sqrt{2A}\rho)/\sqrt{2A}$,
$t^{\se\sm} = \mbox{arccosh}(\sqrt{-2A}\rho)/\sqrt{-2A}$,
$t^{\se\sm} = \sqrt{-2/9K}\rho^{3/2}$, 
$t^{\se\sm} = \rho^2/2(2R - {\cal D}_{\sT\so\st})$ and
$t^{\se\sm} =\sqrt{\rho^2 - {\cal D}_{\sT\so\st}}$.  
Note that I) all these expressions are globally monotonic except in the first `Milne in AdS' case, for 
which one can only have monotonicity for an epoch. 
II) These expressions (and further examples in \cite{Cones}) are invertible to
$\rho = \rho(t^{\se\sm})$. 
Thus one can then replace $\rho$ dependence in the subsequent $l$-equations by $t^{\se\sm}$ dependence: 
eq (5)'s $\hat{H}_{l}$ can be written as $\hat{H}_{l}(\rho, S^{u}) = 
\hat{H}_{l}(t^{\se\sm}, S^{u}) = \hbar^2D_{S}^2/\rho(t^{\se\sm})^2
\hbar^2 + J(\rho(t^{\se\sm}), S^{u})$.

This simplifies further if one uses instead \cite{Cones}
\beq
t^{\sr\se\sc} = t^{\sr\se\sc}(0) + \int\d t^{\se\sm}/\rho(t^{\se\sm})^2 \mbox{ } 
\eeq 
(`rec' stands for `rectified').
For the above five examples, this takes the form
$t^{\sr\se\sc} = \mbox{const} - \mbox{cot}(  \sqrt{2A}t^{\se\sm}  )/(  \sqrt{2A})^{3/2}$,    
$t^{\sr\se\sc} = \mbox{const} + \mbox{tanh}(\sqrt{-2A}t^{\se\sm})/(-2A)^{3/2}$,  
$t^{\sr\se\sc} = \mbox{const} - ({-2}/{K})^{2/3}/{{3t^{\se\sm}}^{1/3}}$, 
$t^{\sr\se\sc} =  \mbox{const} + \mbox{ln}\,t^{\te\tm}/{2\sqrt{2R - {\cal D}_{\sT\so\st}}}$ and  

\noindent $t^{\sr\se\sc} =   \mbox{arctan}({t^{\se\sm}}/({\cal D}_{\sT\so\st} - 2R))/({\cal D}_{\sT\so\st} - 2R) + \mbox{const}  $.  
This is also invertible for the current paper's specific examples and with no loss in monotonicity. 
By this simplification, the time-dependent Schr\"{o}dinger equation is 
\beq
i\hbar\pa_{\st^{\tr\te\tc}}|\chi\rangle = -(\hbar^2/2)D_{S}^2|\chi\rangle + 
\widetilde{J}(t^{\sr\se\sc}, S^{u})|\chi\rangle \mbox{ } \mbox{ where } \mbox{ }  
\widetilde{J} = \rho(t^{\sr\se\sc})^2 J \mbox{ } ,  
\eeq
which can be viewed as a $t^{\sr\se\sc}$-dependent perturbation of what is, in the $N$-stop metroland 
case, a well-known $t^{\sr\se\sc}$-dependent Schr\"{o}dinger equation.  
This is the usual Schr\"{o}dinger equation  on the circle/sphere/hypersphere).
\cite{MGM} contains the 3-stop metroland case of this for HO-like potentials, while this and more 
complicated cases are further studied in \cite{SemiclIII}, in which the back-reaction of   
the $l$-subsystem on the $h$ subsystem is modelled via obtaining tractable systems of perturbative 
equations.  
I will subsequently use such RPM models to also investigate whether the following terms habitually neglected 
in the semiclassical quantum cosmology literature is in fact questionable.
1) Higher time derivative terms (the neglect of higher derivatives can be unreasonable e.g. in fluid dynamics).  
2) Averaged/expectation terms (the neglect of which {\sl is} unreasonable in Molecular Physics, where one 
requires, rather, the Hartree--Fock self-consistent approach that keeps such terms.)

Throughout the above, useful checks of the semiclassical approach's assumptions and approximations 
follow from RPM's having some \cite{08II, AF, +tri, MGM, 08III, SemiclIII} examples that are 
ulteriorly exactly soluble.  
(I.e., exactly soluble by means outside those that are usually available for specific toy models of the 
semiclassical approach, which are, in particular, seldom available in minisuperspace.)

All-in-all, RPM's may be viewed as valuable toy models of midisuperspace Quantum Cosmology models that 
consider the origin of structure formation in the universe. 
(An example of a such is the Halliwell--Hawking set-up \cite{HallHaw} to how Quantum Cosmology may 
seed galaxy formation and CMB inhomogeneities.)
Thus RPM's are valuable conceptually and as regards testing whether we should be {\sl qualitatively} 
confident in the assumptions and approximations made in schemes such as \cite{HallHaw}.  
[0), 1) and 2) above are some of the reasons to wish for such qualitative checks in analytically 
tractable toy models such as RPM's.]

\subsection{Hidden dilational time for $N$-stop metrolands}

This is a further problem of time application of scaled RPM's, developed further in \cite{FileR}.
Scaled RPM possesses an Euler time \cite{06II, SemiclI} analogous to York time. 
In fact, this analogy is not exact, highlighting how in each case there is a multiplicity of scale 
variables whose canonical conjugates can then serve as dilational hidden times \cite{FileR}.  
For RPM's and minisuperspace, the analogue of the Lichnerowicz--York equation \cite{York73}, to be 
solved to isolate the true Hamiltonian $\fH_{\st\sr\su\se}$, is merely algebraic.
How tractable it is varies with choice of scale variable. 
This study reveals that the following can be generalized. 
1) The constant mean curvature lapse-fixing equation (of theoretical numerical relativity \cite{BS03} as 
well as of the hidden York time approach \cite{K92, I93}).  
2) The Lagrange--Jacobi equation for particle mechanics (familiar from Celestial Mechanics). 
Moreover, this study reveals that the generalizations of these two important equations from widely 
different branches of physics are in fact closely inter-related \cite{FileR}.

\subsection{Timeless approach applications of $N$-stop metrolands}

As regards the na\"{\i}ve Schr\"{o}dinger interpretation \cite{HP86,UW89}, \cite{AF, +tri, FileR}'s 
questions concerning shapes, and evaluations of answers to these questions, apply again in the scaled 
case by the shape--scale split. 
One can now also investigate questions about the scale of the model universe, e.g. what is the 
probability P(moment of inertia of the model universe lies between 0 and $I$). 
E.g. for the ground state of the special multi-HO 3-stop metroland model, 

\beq
\mbox{P(moment of inertia of the model universe lies between 0 and $I$)} \propto 1 - \mbox{exp}(-I/I_{\sH\sO}) \mbox{ } . 
\eeq
Thus it tends to 0 for $I << I_{\sH\sO}$ (the characteristic value of the moment of inertia for this particular problem) 
and to 1 for $I >> I_{\sH\sO}$.

As regards setting up records theories for $N$-stop metroland,   
1) notions of localization in space were considered in \cite{AF}.  
2) notions of localization in configuration space include \cite{Records} ones based on the reduced 
configuration space metric. 
(This is useable here even though it does not extend to geometrodynamics as a bona fide notion of 
distance \cite{NOD}, so that investigating alternatives is also desirable \cite{NOD}).  
3) One can build up notions of quantum entropy/information from wavefunctions such as those provided in 
this paper.  
Moreover, no standard type of ensemble would appear to be appropriate. 
For, RPM's are whole-universe models and thus of fixed total energy, but with variable numbers of 
particles (due to coalescence), which does not fit the description of any of the microcanonical, 
canonical or grand ensembles.  
Due to that and low particle number issues, I consider that building entropies (e.g. along the lines of the 
von Neumann entropy) from the wavefunctions currently lacks justification. 
(Low particle number issues can be dealt with by considering the present paper's work with $N$ large, 
%
%
while \cite{H99}'s study demonstrates how more finite systems do still have nontrivial, if imperfect, 
notions of information/records.)

\subsection{Histories and the joint Semiclassical/Histories/Records approach}

3-stop metroland is straightforward to present in a histories theory formulation since it is 
unconstrained and geometrically simple.
For the more complicated triangleland case of this, see \cite{08III, FileR}.

Thus, combining with \cite{SemiclIII}'s semiclassical approach developments, I have shown that 
one can get far with each of the three Problem of Time strategies that I am next proposing to combine 
(histories, records and the semiclassical approach).
The next step of this program will involve producing more complicated and genuinely closed-universe versions 
of Halliwell's work on various partial combinations of these three approaches \cite{H99, H03, H09} 
(see \cite{ASharp} for a more detailed outline of this next step.)

\subsection{Applications to other fundamental issues in Quantum Cosmology}

Further foundational issues in Quantum Cosmology with at least qualitative counterparts in     
RPM toy models are as follows.  
Does structure formation in the universe have a quantum-mechanical origin?  
In GR, studying this requires midisuperspace or at least inhomogeneous perturbations about 
minisuperspace \cite{HallHaw}, which are of great difficulty.     
There are also a number of difficulties associated with 1) closed system physics and observables, 
speculations on initial conditions. 
2) the meaning and form of $\Psi$ (e.g. the discussion of uniformity 
below) and the origin of the arrow of time \cite{HH83, EOT, H03, Rovellibook}.

Uniformity is of widespread interest in cosmology.  
As well as applying to good approximation to the present distribution of galaxies and to the CMB, 
this relevant to 1) whether there was a considerably more uniform initial state \cite{Penrose}.  
2) related issues of uniformizing process and how the small perturbations observed today were seeded.
I note that uniformity notions are the same for scaled and scalefree RPM's.  
Thus this paper inherits these from \cite{AF,+tri}, while it is the present paper's scaled RPM's that 
constitute a superior arena for modelling GR cosmology.  
This paper's study of the 3-stop metroland case is a useful `precursor' for the more complex of notions 
of uniformity in bigger models'.   
In the equal-mass case of 3-stop metroland, there is one notion of merger and this coincides with 
equally-separated-out particles.  
For triangleland with equal masses there are the equilateral triangle states at the poles that are the 
most uniform states.  
There are two ways of departing from this uniformity. 
1) Tending towards collinearity.  
2) Tending away from the great circles of isoscelesness that join the equilateral configurations to the 
3 M-points on the equator of collinearity (that coincide with the 3-stop metroland notion).  
4-stop metroland then has numerous notions of merger as discussed in \cite{FileR}, one point among 
which additionally involves the four equal masses being equally spaced out.
Uniformity also makes a good focus for further investigations using the na\"{\i}ve Schr\"{o}dinger 
interpretation and other timeless approaches.

Finally, there is also a robustness issue: does ignoring some degrees of freedom substantially change a 
Quantum Cosmology? 
Kucha\v{r} and Ryan's \cite{KR89} criteria for such an investigation are exact solutions to 
approximations sitting inside exactly soluble models.  
Scaled RPM's alongside the Cosmology-RPM analogy give further such examples.  
Furthermore, exact solvability of nontrivial models within models is likely confined to 
the present paper's 1-$d$ RPM setting \cite{FileR}.  

\mbox{ }

\noindent{\bf Acknowledgements}  

\mbox{ } 

\noindent My wife Claire and my friends Alicia, Amy, Joshua, Emily, Coryan, Sophie, Sophie, Sally and Tea  
for their support.  
Dr Julian Barbour, 
Miss Anne Franzen, 
Mr Sean Gryb, 
Mr Henrique Gomes and 
Prof Karel Kucha\v{r} for discussions.
Dr Merced Montesinos, Mr Simeon Bird, Prof Jonathan Halliwell and the anonymous Referees for comments.
Professors Malcolm MacCallum, Reza Tavakol, Gary Gibbons, Don Page, Jeremy Butterfield, Belen Gavela, 
Enrique Alvarez and Marc Lachi\`{e}ze-Rey and Dr Alexei Grinbaum for help with my career.  

\mbox{ }


\noindent{\bf \large Appendix: the Hamiltonian RPM--Cosmology analogy}  

\mbox{ }

\noindent
Because of the 3-metric nature of the variables, for e.g. isotropic minisuperspace with q scalar fields 
$\chi_A$, $p_a \propto - a\dot{a}$ and $p_{\chi_A} \propto a^3\dot{\chi}$, giving, in Hamiltonian form, 
\beq
\frac{1}{a^4}
\left(
p_a\mbox{}^2 - \frac{p_{\chi}\mbox{}^2}{a^2}
\right) 
\propto -\frac{k}{a^2} + \frac{2GM_{\sd\su\sss\st}}{a^3} + \frac{2GM_{\sr\sa\sd}}{a^4} + 
\frac{\Lambda}{3} \mbox{ } , 
\eeq
which has a different correspondence from the Lagrangian one in the main text.  
Now curvature corresponds to HO's, cosmological constant to fourth-order potentials, dust to linear 
potentials and radiation to constant potential.  
Moreover, note that the case of constant plus HO-type potential that is straightforward to study in 
mechanics continues to be relevant in this analogy.  
This analogy is useful for quantum scale equations, though if these are treated semiclassically as is my 
main intention, the analogy reverts to that in the main text.

However, if one takes the quantum scale equation, one gets a direct analogy. 
[This ignores the disparity in sign between shape and scalar field kinetic terms.] 
It is for p(q -- 1) = 2(f -- 1) for p the spatial dimension of the GR (usually 3), f the dimension of 
the RPM configuration space and q the number of minimally-coupled scalar fields (mcsf's).  
Thus e.g. 3-stop metroland is analogous to 1 mcsf in any dimension.   
4-stop metroland has no such analogies for nontrivial geometrodynamics (in the sense of p $\geq 3$ being required to have local 
degrees of freedom). 
5-stop metroland is analogous to 1 mcsf in 5-$d$ spacetime.  
Triangleland likewise has no such analogy, while quadrilateralland corresponds to e.g. 3 mcsf's in 
4-$d$ spacetime.
Examples of analogies with 11-$d$ spacetimes include (beyond the ubiquitous 3-stop example) 
8-stop metroland and pentagonland; each of these has 2 mcsf's.
Whether there are further such analogies with nonminimally coupled scalar fields, anisotropies and 
inhomogeneities remains work in progress.  
One early widespread indication is that it is positive potential powers that are common (i.e. without 
the abysses and instabilities of the $1/|{\bf r}^{ab}|$ potentials) in Quantum Cosmology. 
This is one further reason why I mostly concentrate on positive-power models in the present paper.


\end{document}